\documentclass[twocolumn,showpacs,superscriptaddress,preprintnumbers,amsmath,amssymb,prc]{revtex4-2}

\usepackage{graphicx}
\usepackage{dcolumn}
\usepackage{bm}
\usepackage{float}
\usepackage{color}
\usepackage{ulem}
\usepackage{CJK}
\usepackage[table,usenames,dvipsnames]{xcolor}
\definecolor{linkcolor}{rgb}{0.6,0,0}
\definecolor{citecolor}{rgb}{0,0,0.75}
\definecolor{urlcolor}{rgb}{0.12,0.46,0.7}
\usepackage[breaklinks, colorlinks, urlcolor=urlcolor, linkcolor=linkcolor,citecolor=citecolor,pdfencoding=auto]{hyperref}
\hypersetup{linktocpage}

\usepackage{tikz,xcolor,hyperref}
\definecolor{lime}{HTML}{A6CE39}
\DeclareRobustCommand{\orcidicon}{
\begin{tikzpicture}
\draw[lime, fill=lime] (0,0) 
circle [radius=0.16] 
node[white] {{\fontfamily{qag}\selectfont \tiny ID}};
\draw[white, fill=white] (-0.0625,0.095) 
circle [radius=0.007];
\end{tikzpicture}
\hspace{-2mm}
}
\foreach \x in {A, ..., Z}{%
\expandafter\xdef\csname orcid\x\endcsname{\noexpand\href{https://orcid.org/\csname orcidauthor\x\endcsname}{\noexpand\orcidicon}}
}

\begin{document}
\begin{CJK*}{UTF8}{gbsn}

\title{AdS black hole thermodynamics and microstructures from $f(Q)$ gravitation}
\author{Oleksii Sokoliuk\orcidA{}}
\email{oleksii.sokoliuk@mao.kiev.ua}
\affiliation{Main Astronomical Observatory of the National Academy of Sciences of Ukraine, 27 Akademik Zabolotny St., Kyiv, 03143, Ukraine}
\affiliation{Astronomical Observatory, Taras Shevchenko National University of Kyiv,
3 Observatorna St., 04053 Kyiv, Ukraine}

\author{Sneha Pradhan\orcidB{}}
\email{snehapradhan2211@gmail.com}
    \affiliation{Department of Mathematics, Birla Institute of Technology and
Science-Pilani, Hyderabad Campus, Hyderabad-500078, India}
\author{Alexander Baransky\orcidC{} }
\email{abaransky@ukr.net}
\affiliation{Astronomical Observatory, Taras Shevchenko National University of Kyiv,
3 Observatorna St., 04053 Kyiv, Ukraine}
\author{P.K. Sahoo\orcidD{}}
\email{Corresponding author: pksahoo@hyderabad.bits-pilani.ac.in}
\affiliation{Department of Mathematics, Birla Institute of Technology and
Science-Pilani, Hyderabad Campus, Hyderabad-500078, India}

\date{\today}

\begin{abstract}
The significant properties and phase transition of charged Anti-de Sitter (AdS) black holes have been extensively studied in a variety of modified theories of gravity in the presence of numerous matter fields. The goal of our current research is to investigate the AdS black hole's thermodynamics under the impact of $f(Q)$ gravity. Additionally, this paper explores the black hole's local stability and phase structure under the relevant gravity. Besides, we use Ruppeiner geometry to look into the AdS black hole's microscopic structure. We have numerically computed the Ricci curvature scalar $R$ to explain the interactions between the AdS black hole's microscopic particles under the influence of $f(Q)$ gravity.
\end{abstract}

\maketitle

\section{Introduction}\label{sec:1}
Black-hole solutions come in a variety of forms, including Schwarzschild, Kerr, and BTZ black holes, etc. A black hole solution in general relativity, known as an anti-de Sitter (AdS) black hole depicts an isolated enormous entity with a negative cosmological constant ($\Lambda<0$), whereas the solutions represent the de Sitter space when the cosmological constant is positive ($\Lambda>0$). The groundbreaking work of Bardeen, Carter, and Hawking, \cite{bardeen1973four,hawking1975particle} can be regarded as the pioneer of the research on black hole thermodynamics. The remarkable publication by Hawking and Page \cite{hawking1983thermodynamics}, which established the presence of a specific phase transition in the phase space of the (non-rotating, uncharged) Schwarzschild-AdS black hole, marked the beginning of the history of the study of the thermodynamic characteristics of AdS black holes. Since then, a variety of backdrops with increasingly intricate backgrounds have been studied in the phase transitions and critical phenomena of the AdS black hole.
The thermodynamic properties of AdS black holes, which are fundamental to the study of quantum gravity, have attracted the attention of numerous scientists. It was discovered during that time that AdS black hole space-time geometry can't just be attributed to ordinary thermodynamical properties like pressure, entropy, or temperature but they have also been demonstrated to have rich phase configuration and to admit critical behaviors \cite{eom2022hawking}.

The study of black hole thermodynamics could be separated into two main phases, namely the classical and the extended phase space (EPS) stages, for both historical and scientific reasons. Conventional formalism primarily focuses on the development of thermodynamic relations and the assessment of thermodynamic variables like pressure, temperature, etc. in distinct black hole environments. It is worth to notice that thermodynamics of different types of black holes have been extensively investigated by several researchers from various perspectives \cite{pourhassan2020pv,Ghosh:2020tgy,javed2023weak,sood2022thermodynamic,Ovgun:2017bgx,Pourhassan:2018chj}. Black holes in AdS space have very different thermodynamic characteristics from those in de Sitter space or asymptotically flat spacetimes. Large black holes are thermodynamically stable in AdS space, whereas small ones are unstable. In addition, there is no black hole solution in AdS space below a specific temperature, and in AdS space, a phenomenon known as the Hawking-Page phase transition can occur between stable large black holes and thermal gas \cite{hawking1983thermodynamics}.

By incorporating an additional ($P$,$V$) pair of variables, where $P$ are connected to the cosmological constant by the relation $P=-\frac{\Lambda}{8\pi G}$, the extended phase space formalism can be introduced (it was firstly proposed in the work of \cite{kastor2009enthalpy}). That inaugurated a new era for the thermodynamics of black holes in AdS spacetime. In precise, if we use $G=c=k=1$ then in the case of a 4-dimensional asymptotically AdS black hole system the pressure is identified via the following relation: $P=-\frac{\Lambda}{8\pi }$. In general, the corresponding black hole thermodynamic volume is given by $V=\frac{4}{3} \pi r_{+}^3$ (which only applies to spherically symmetric solutions), where $r_{+}$ is referred to as the radius of the black hole event horizon formulated in terms of the Boyer–Lindquist radial coordinate. This provided the inspiration for several more works, namely \cite{dolan2011pressure,kubizvnak2012p,cai2013pv} (and references therein). These developments are focused mostly on the thermodynamic properties, particularly on criticalities within $P-V$ phase plane. An interesting phase transition has been observed between the large black hole and a small black hole in a canonical ensemble of black holes with a fixed charge \cite{chamblin1999charged}. With regard to charge $\mathcal{Q}$ and temperature $T$ of a black hole, the phase diagram contains a critical point. The charged black holes critical behavior and phase transitions are remarkably similar to those of the Van der Waals liquid-gas system, whereas the cosmological constant is a parameter for a particular theory in the first law of black hole thermodynamics, while the mass $M$, angular momentum $J$, and charge $\mathcal{Q}$ are conserved charges also in some cases the cosmological constant should be viewed as a variable for a number of physical reasons \cite{chamblin1999charged}. There are many groundbreaking works of the great scientist Jacob D. Bekenstein on the field of black hole thermodynamics, which sheds light on this subject for the young researcher. In the work \cite{B1}, the author has introduced the concept of black hole entropy, which he has associated with the surface area of the black hole. He suggested that the appropriate generalization of the second law for a region containing a black hole is that the black-hole entropy plus the common entropy in the black-hole exterior never decreases. After that renowned work, the author proved the applicability of the second law of thermodynamics for the infall of an entropy-bearing system into a much larger and more powerful generic stationary black hole \cite{B2}.

The study of the charged AdS black hole's critical behavior in the extended phase space has recently made the comparison between a charged black hole in AdS space and a Van der Waals system possible and such comparison were published in a several papers: \cite{chamblin1999charged,shen2007thermodynamic,alfaia2022central}. The $P-V$ diagram of the charged AdS black hole is exactly the same as the one for the Van der Waals liquid-gas system, according to the authors of a reference \cite{kubizvnak2012p}, who started the investigation of the $P-V$ critical behavior of the black hole in the extended phase space. The critical exponents for the charged AdS black hole phase transition are identical to those for the Van der Waals system and therefore, comparison between the charged AdS black hole and the Van der Waals system completes itself. But remind that though the $P-V$ diagram of the Van der Waals system is equivalent to the graph of $\mathcal{Q}-\Phi$ in charged AdS black hole where $P$ and $V$ are the pressure and volume of the Van der Waals system, respectively, and $\Phi$ is the chemical potential conjugate to the black hole's charge $\mathcal{Q}$, the comparison is controversial as $\Phi$ is an intensive quantity and the charge $\mathcal{Q}$ referred to as an extensive one in black hole thermodynamics, whereas $V$ is an extensive quantity and $P$ is an intensive one in the Van der Waals system \cite{shen2007thermodynamic}. Besides these topics, an important area of research has been figuring out the 
 microscopic structure of black holes. Ruppeiner geometry is the most successful tool to study the microscopic structure of a charged AdS black hole. One may follow the work based on the Ruppeiner geometry, which explores the microstructures of a black hole solution in \cite{wei2022microstructure,wei2019ruppeiner}. The most effective method for examining a charged AdS black hole's microstructures is the so-called Ruppeiner geometry. Each geometrical quantity, such as the line element $dl^2$ and the Ricci scalar curvature $R$, has a physical meaning that encodes the microscopic structure of the AdS black hole system.

 Although it is well known that Einstein's general relativity is an outstanding tool for discovering many hidden mysteries of nature, the theory has been theoretically challenged by specific observable evidence of the expanding universe and the presence of dark matter. As a result, several modified theories of gravity have been proposed time by time. There have been a certain number of works on AdS black holes in modified theories of gravity. For example in the paper \cite{nashed2021new} the author has investigated the physical properties and stability of AdS black hole within $f(R)$ theory. They have derived a stable black hole solution as well as shown that the derived solution must satisfy the first law of thermodynamics. Moreover, Ping Li and his group \cite{li2018thermodynamics} studied widely the thermodynamics of charged AdS black holes in Rainbow gravity. They discovered that the mass of the test particle has a unique value that would cause the black hole to experience zero temperature and divergent heat capacity with a fixed charge. Other researchers in the paper \cite{heisenberg2018hairy} have studied the properties of black holes on a static and spherically symmetric background under the $U(1)$ gauge-invariant scalar-vector-tensor theories. In the research work \cite{heisenberg2017hairy} authors have presented a set of exact black hole solutions with the background of the static spherically symmetric case, in the second-order generalized Proca theories with derivative vector-field interactions coupled to gravity.  Most importantly in the work \cite{d2022black} for $f(Q)$ gravity, the authors have methodically developed and investigated symmetry-reduced field equations, and as well as for $f(T)$ gravity, they have sketched out how to take a similar approach.
Here in this paper, we will study the thermodynamic behavior and phase transition of the black hole in AdS spacetime under $f(Q)$ theory of gravity in which cosmological constant is considered as dynamical pressure that the system experiences and the conjugate quantity of cosmological constant are the black holes' thermodynamic volume. 

The paper is organized as follows: In Section (\ref{sec:1}) we provide the literature survey and introduction to the topic of AdS black holes and their thermodynamical features. In the next section, (\ref{sec:2}) we present the modified gravity framework, namely symmetric teleparallel gravitation $f(Q)$, and derive the Equations of Motion for such gravity theory. On the other hand, in Section (\ref{sec:3}) we finally present the metric tensor for perturbed up to the second order of $\alpha$ AdS black hole solution within quadratic $f(Q)$ gravitation that mimic Starobinsky $f(R)$ solution. In the following subsections, we derive such thermodynamical quantities as Hawking temperature, heat capacity, and Helmholtz free energy. In Section (\ref{sec:4}) we derive Ruppeiner curvature and examine the microstructure properties of our BH solution. Finally, in the last section, namely Section (\ref{sec:5}) we provide the concluding remarks on the key topics of our study.

\section{$f(Q)$ gravitation framework}\label{sec:2}
In the current section, we are going to present the foundations of $f(Q)$ gravity formalism. Firstly, it is worth saying that we will work under the assumption that we live on the differentiable Lorentzian manifold $\mathcal{M}$. Such manifold and its dynamics could be generally described by the metric tensor $g_{\mu\nu}$, its determinant $g=\det g_{\mu\nu}$ and metric-affine connection $\Gamma$ (that respectively matches spacetime basis with its derivatives, is not metric compatible for the sake of non-metricity existence). Therefore, one could formulate the general affine connection in one form \cite{ortin2015}:
\begin{equation}
\Gamma^{\alpha}_{\,\,\,\,\beta}=w^{\alpha}_{\,\,\,\,\beta}+K^{\alpha}_{\,\,\,\,\beta}+L^{\alpha}_{\,\,\,\,\beta}.
\end{equation}
It respectively consists of Levi-Civita, contortion, and disformation one forms. The equation above could be rewritten in the more recognizable form below:
\begin{equation}
\Gamma^{\alpha}_{\,\,\,\,\mu\nu}=\gamma^{\alpha}_{\,\,\,\,\mu\nu}+K^{\alpha}_{\,\,\,\,\mu\nu}+L^{\alpha}_{\,\,\,\,\mu\nu}.
    \label{eq:2}
\end{equation}
In the equation above, $\gamma^{\alpha}_{\mu\nu}$ is the regular Levi-Civita metric-compatible affine connection that appears in the general theory of relativity (this case could be easily recreated with the vanishing torsion and non-metricity). On the other hand, in order to recreate symmetric teleparallel gravitation, one could assume that the only non-metricity term fully describes the gravitation. A metric-compatible, curvature-free connection with torsion gives rise to the theory called Teleparallel Equivalent of GR(TEGR) while the vanishing torsion and curvature give Symmetric Teleparallel Equivalent of GR (STEGR). In STEGR, the non-metricity tensor is left with : 
\begin{equation}
    \boldmath{Q_{\alpha\mu\nu}=\nabla_{\alpha}g_{\mu \nu}=-L^{\beta}_{\,\,\,\,\alpha\mu}g_{\beta\nu}-L^{\beta}_{\,\,\,\,\alpha\nu}g_{\beta\mu},}
\end{equation}
which could be properly defined by the expression provided below:
\begin{equation}
    L^\alpha_{\,\,\,\,\mu\nu}=\frac{1}{2}Q^{\alpha}_{\,\,\,\,\mu\nu}-Q_{(\mu\nu)}^{\,\,\,\,\,\,\,\, \alpha}.
    \label{eq:77}
\end{equation}
The foundation of the symmetric teleparallelism is described generally by the non-metricity scalar:
\begin{equation}
    Q=-P^{\alpha\mu\nu}Q_{\alpha\mu\nu}.
\end{equation}
The non-metricity conjugate is defined as,
\begin{equation}
\boldmath{P_{\,\,\,\,\mu\nu}^{\alpha}=-\frac{1}{2}L^\alpha_{\,\,\,\,\mu\nu}+\frac{1}{4}\left(Q^{\alpha}-\tilde{Q^{\alpha}}\right)g_{\mu\nu}-\frac{1}{4}\delta^{\alpha}_{\,\,(\mu\,} Q_{\nu)},}
\end{equation}
where
\begin{equation}
   \boldmath{ Q_{\alpha}=Q_{\alpha \,\,\,\, \nu}^{\,\,\,\, \nu},  \quad \overline{Q}_\alpha=Q^\mu_{\,\,\,\,\, \alpha\mu},}
\end{equation}
which is included in the Einstein-Hilbert (further - EH) action of the theory \cite{Xu2019}:
\begin{equation}
    \mathcal{S}[g,\Gamma,\Psi_i]=\int d^4x \sqrt{-g}f(Q)+\mathcal{S}_{\mathrm{M}}[g,\Gamma,\Psi_i].
    \label{eq:10}
\end{equation}
In the equation above $d^4x=dtdrd\theta d\phi$, $f(Q)$ is the arbitrary function of the non-metricity scalar that defines the corresponding theory of $f(Q)$ gravity. Besides, $\mathcal{S}_{\mathrm{M}}[g,\Gamma,\Psi_i]$ denotes the contribution of various perfect and non-perfect fluid matter fields $\Psi_i$ (both minimally and non-minimally coupled to gravity) to the total EH action. Remarkably, one could get the proper equations of motion for any theory via the principle of the least action $\delta \mathcal{S}=0$, so that by taking the variational derivative of action with respect to the metric tensor inverse $g^{\mu\nu}$ we get:
\begin{equation}
\begin{gathered}
\frac{2}{\sqrt{-g}}\nabla_\gamma\left(\sqrt{-g}\,f_Q\,P^\gamma\;_{\mu\nu}\right)+\frac{1}{2}g_{\mu\nu}f \\
+f_Q\left(P_{\mu\gamma i}\,Q_\nu\;^{\gamma i}-2\,Q_{\gamma i \mu}\,P^{\gamma i}\;_\nu\right)=-T_{\mu\nu},
\end{gathered}
\label{eq:11}
\end{equation}
Here we assume that $f_Q=df/dQ$, $\nabla_\mu$ is a covariant derivative that takes into account spacetime curvature and that $T_{\mu\nu}$ is the generalized stress-energy-momentum tensor for any matter content. Without the loss of generality, we could define this tensor as a variation of matter content Lagrangian density w.r.t. $g^{\mu\nu}$:
\begin{equation}
    T_{\mu\nu}=-\frac{2}{\sqrt{-g}}\frac{\delta(\sqrt{-g} \mathcal{L}_{\mathrm{M}})}{\delta g^{\mu\nu}}.
\end{equation}
Therefore, since we already briefly discussed the framework of modified gravity that we will be working within, we could proceed further and derive the hydrostatic equilibrium equations.
\section{AdS black hole solution in $f(Q)$ cosmologies}\label{sec:3}
In this subsection, we are correspondingly going to obtain field equations and other important quantities for modified $f(Q)$ gravitation with the generalized metric tensor of $\mathrm{sig}(g)=(-,+,+,+)$.
\begin{equation}
    ds^2 = g_{tt}dt^2+g_{rr}dr^2+r^2(d\theta^2+\sin^2\theta d\phi^2).
\end{equation}
In order to obtain the approximate form of $g_{tt}$ and $g_{rr}$, one needs to assume the function $f(Q)$. In our case, we choose $f(Q)=Q+\alpha Q^2-2\Lambda$, which recreates the Starobinsky-like quadratic model (for more information on the subject, refer to \cite{Starobinsky:1980te}) with asymptotically AdS behavior and one degree of freedom, the namely free parameter $\alpha$. In the $f(Q)$ gravitation theory, assuming electro-vacuum energy-momentum tensor \cite{PhysRevD.105.024042}:
\begin{equation}
    T^\mu_{\nu}=\mathrm{diag}\bigg(\Lambda+\frac{\mathcal{Q}^2}{r^4},\Lambda+\frac{\mathcal{Q}^2}{r^4},\Lambda-\frac{\mathcal{Q}^2}{r^4},\Lambda-\frac{\mathcal{Q}^2}{r^4}\bigg),
\end{equation}
and perturbed metric, affine connection (expanded around $\alpha$)
\begin{align}
    &g_{tt} = g_{tt}^{(0)}+\alpha g_{tt}^{(1)}+\alpha^2 g_{tt}^{(2)},\\
    &g_{rr} = g_{rr}^{(0)}+\alpha g_{rr}^{(1)}+\alpha^2 g_{rr}^{(2)},\\
    &\Gamma^r_{\theta\theta} = -r+\alpha \gamma^{(1)}+\alpha^2 \gamma^{(2)}.
\end{align}
Where $\mathcal{Q}$ is the black hole charge, that arises from the Maxwell field contribution to the Einstein-Hilbert action integral and $\Lambda$ is the cosmological constant or $\Lambda$ term. Zero order forms of $g_{tt}$ and $g_{rr}$ corresponds to the GR RN-AdS black hole solution :
\begin{equation}
    g_{tt}^{(0)}=-\bigg(1-\frac{2M}{r}+\frac{\mathcal{Q}^2}{r^2}-\frac{\Lambda}{3}r^3\bigg),\quad g_{rr}^{(0)}=-\frac{1}{g_{tt}^{(0)}}.
\end{equation}
Additionally, we assume that the black hole charge vanishes so that we have the solution for the AdS black hole \cite{PhysRevD.105.024042}:
\begin{equation}
\begin{gathered}
    g_{tt}= -1+\frac{2M_\text{ren}}{r}+\frac{\Lambda_\text{ren}}{3}r^2+\alpha^2\mu\left(\frac{2M_\text{ren}}{r}+\frac{\Lambda_\text{ren}}{3}r^2\right)\\
    \times\ln\left(\frac{r^{*2}}{r^2}\left(\frac{2M_\text{ren}}{r}+\frac{\Lambda_\text{ren}}{3}r^2\right)\right),\quad g_{rr}=-\frac{1}{g_{tt}}.
\end{gathered}
\label{eq:19}
\end{equation}
Where for simplicity we defined the set of new variables
\begin{equation}
    2M_\text{ren} =2M+\frac{\alpha(c_2+\alpha c_4)+24\alpha^2M(c_6-c_7)}{9},
\end{equation}
\begin{equation}
    \frac{\Lambda_\text{ren}}{3} =\frac{\Lambda}{3}\left(1+\frac{12\alpha^2(c_6-c_7)}{9}\right),
\end{equation}
\begin{equation}
    \mu = \frac{4c_8}{18M}.
\end{equation}
There are five unknown constants $c_i$, namely constants of integration, that arise from the field equations. For the sake of simplicity, we will ignore $c_4$, which comes from the solving field equation at second order of $\alpha$. Furthermore, we fix $c_8$ to be quite small, so that the logarithmic term in the metric potential could be neglected at the event horizon. With such an assumption, we could control the beyond-GR corrections of metric potential. Now we are left with four degrees of freedom namely $\alpha$, $c_2$, $c_6$ and $c_7$. For the sake of simplicity, it is possible to reduce the dimensions of the system by assuming that $c_6-c_7=c_9$. Finally, in the equations above, $r_*$ is the arbitrary scale that defines at which point the logarithmic term in (\ref{eq:19}) will dominate over the Schwarzschild-AdS contribution. For further investigation, one needs to derive the functional form of the BH mass $M$ by matching $g_{tt}(r_+)=0$ with $r_+$ being the event horizon radius:
\begin{equation}
    M(r_+)=\frac{9 r_+-\alpha  c_2}{24 \alpha ^2
   c_9+18}-\frac{\Lambda  r_+^3}{6}.
\end{equation}
For mass to be positive, our free parameters need to respect the following inequalities (for $c_9<0$):
\begin{equation}
0<\alpha <\frac{1}{2} \sqrt{3}
   \sqrt{-\frac{1}{c_9}} ,
\end{equation}
\begin{equation}
 c_2<\frac{-4
   \alpha ^2 c_9 r^3 \Lambda -3 r^3 \Lambda +9 r}{\alpha
   }.
\end{equation}
On the other hand, for other domains of $\alpha$, we have that
\begin{equation}
\alpha >\frac{1}{2} \sqrt{3}
   \sqrt{-\frac{1}{c_9}} ,
\end{equation}
\begin{equation}
    c_2>\frac{-4
   \alpha ^2 c_9 r^3 \Lambda   -3 r^3 \Lambda +9 r}{\alpha
   }.
\end{equation}
Finally, if one will assume that $c_9\geq 0$, constraints will differ significantly:
\begin{equation}
   \alpha >0\land c_2<\frac{-4
   \alpha ^2 c_9 r^3 \Lambda  -3 r^3 \Lambda +9 r}{\alpha
   }.
\end{equation}
In the further investigation, we will assume that $\Lambda=-\frac{(D-1)(D-2)}{2l^2}=-3l^2<0$ with $l$ being the unitary AdS radius.

\subsection{Derivation of various thermodynamical quantities}

In this subsection, we are going to derive the exact form of various thermodynamical quantities, such as thermodynamical pressure, entropy, heat capacity, and Hawking temperature. We will start with the last quantity Hawking temperature, which is usually defined by the expression \cite{Deng:2017abh}:

\begin{equation}
    T_{\mathrm{H}}=\frac{-g'_{tt}(r_\mathrm{+})}{4\pi}.
    \label{eq:29}
\end{equation}
Here $r_+$ is the outer horizon of $f(Q)$-AdS black hole solution, which is defined as the largest root of $g_{tt}(r)=0$. Following the works of \cite{Cai:1998vy,PhysRevD.65.084014}, one could also define the black hole entropy in terms of mass and Hawking temperature (derived from the first law of thermodynamics):
\begin{equation}
    S=\int \frac{dM}{T_{\mathrm{H}}}=\int_0^{r_+} \frac{1}{T_{\mathrm{H}}}\bigg(\frac{\partial M}{\partial r_+}\bigg)dr_+ .
    \label{eq:30}
\end{equation}

In the current study, in addition to the entropy and temperature we add cosmological constant as a pressure $P$ into our thermodynamical system \cite{Zhang:2021raw}:
\begin{equation}
    P=-\frac{\Lambda}{8\pi}=\frac{3}{8\pi l^2},
    \label{eq:31}
\end{equation}
corresponding conjugate volume element, therefore  (simply the volume of 3-sphere with radius $r_+$):

\begin{equation}
V=\frac{4\pi r_+^3}{3}.
\label{eq:32}
\end{equation}

Finally, with the use of expressions for both entropy and Hawking temperature, one could easily derive the heat capacity at constant pressure, the fundamental quantity of any black hole solution:

\begin{equation*}
C_P = T_{\mathrm{H}}\bigg(\frac{\partial S}{\partial T_\mathrm{H}}\bigg)\biggr \rvert _{P} = T_{\mathrm{H}}\bigg(\frac{\partial S}{\partial r_+}\frac{\partial r_+}{\partial T_{\mathrm{H}}}\bigg)\biggr \rvert _{P}.
\end{equation*}

The equation above can be simplified using the 
expression for black hole entropy from (\ref{eq:30}):

\begin{equation}
 C_{P}=\frac{\partial M}{\partial T_{\mathrm{H}}}=\frac{\partial M}{\partial r_{+}}\frac{\partial r_{+}}{\partial T_{\mathrm{H}}}.
\end{equation}

In general, positive heat capacity could lead to a stable black hole solution. On the other hand, for the case when $C_{P}<0$, the black hole solution is very unstable and could disappear even because of the small exterior perturbation.

\subsection{Results}
\begin{figure*}[!htbp]
    \centering
    \includegraphics[width=\textwidth]{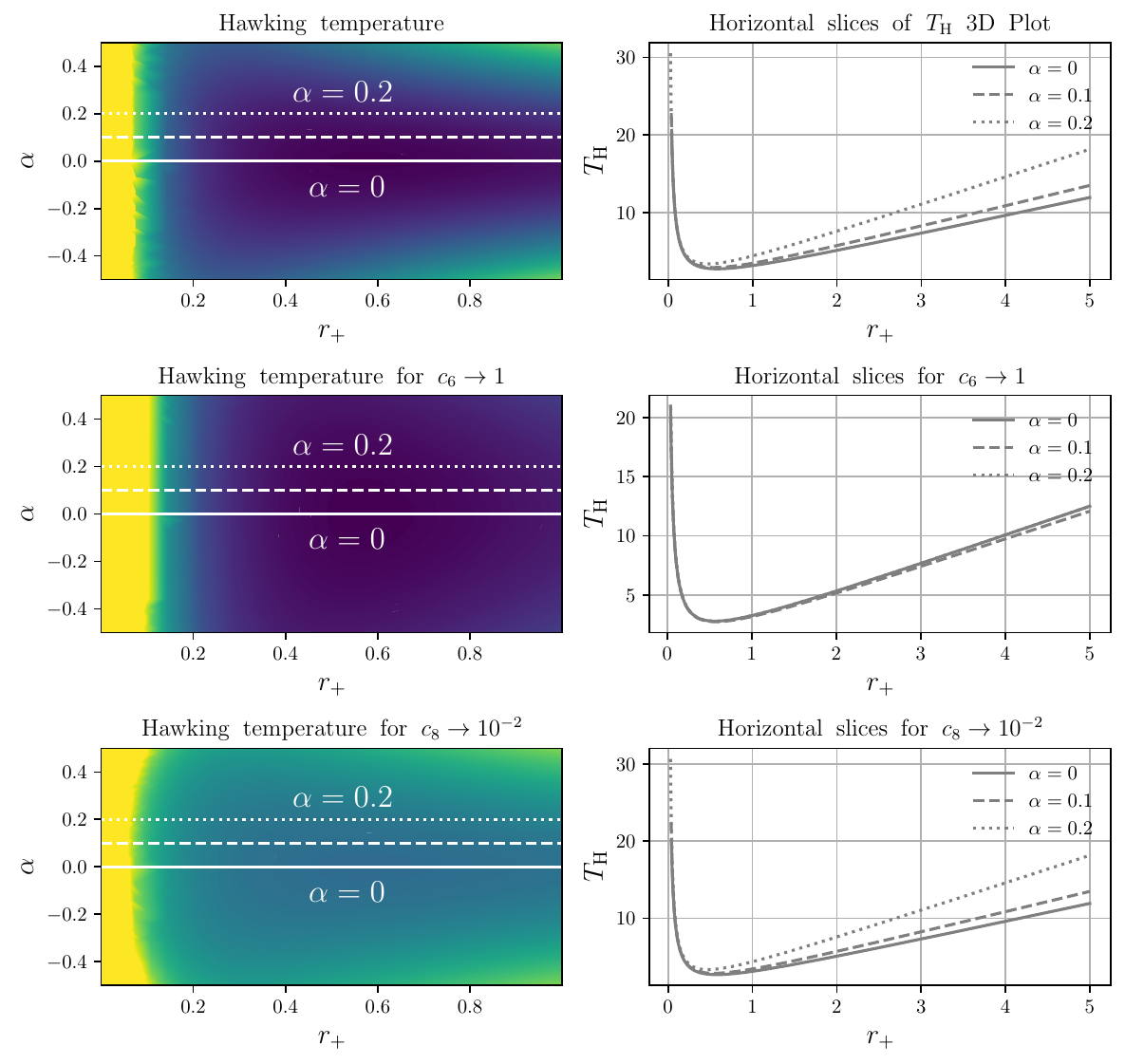}
    \caption{Numerical calculations of Hawking temperature for $f(Q)$-AdS black hole with (\textit{first row}) $\Lambda=-3$, $c_2=10^{-5}$, $c_9=9.9$, $c_8=10^{-5}$ and scale is chosen to be $r^*=10\gg r_+$, (\textit{second row}) we change $c_9$ and assume that it is $0.9$, all other constants remain the same, (\textit{third row}) we change $c_8\to 10^{-2}$, all other constants remain the same}
    \label{fig:1}
\end{figure*}

At first, we are going to numerically investigate such fundamental quantity as Hawking temperature, which is defined as a temperature of Hawking radiation measured by an asymptotic observer located at $r\to\infty$. With the use of various numerical solvers, we obtained the solutions for (\ref{eq:29}) graphically and located them in Figure (\ref{fig:1}). For the sake of completeness, we varied event horizon radii $r_+$, $\alpha$ and additionally $c_8$, $c_9$. As it turned out, temperature and black hole behavior does not really change with the variation of $c_9$ within the positive domain of its values. Moreover, to obtain solutions for Hawking temperature that does not diverge at $r_+\gg 0$ and are positive, one needs to assume that $c_2\ll 1$. Finally, from the numerical analysis of Hawking's temperature, we found out that it generally rises with the bigger values of $\alpha$, and gets smaller with $c_9<0$. Moreover, if values of $c_8$ will grow, $T_\mathrm{H}\to\infty$, and therefore black hole will evaporate quicker.

\begin{figure*}[!htbp]
    \centering
    \includegraphics[width=\textwidth]{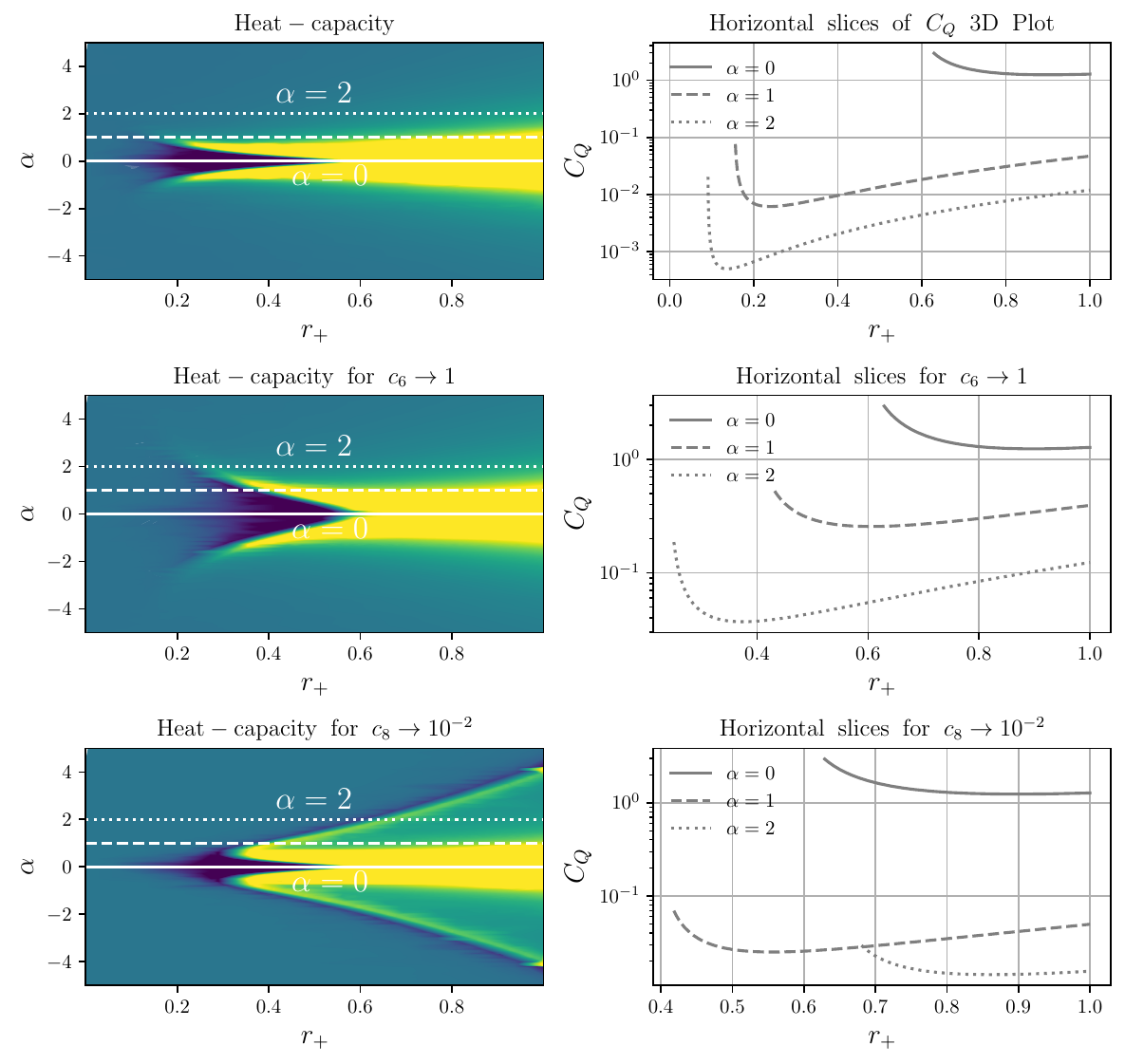}
    \caption{Numerical calculations of heat-capacity for $f(Q)$-AdS black hole with (\textit{first row}) $\Lambda=-3$, $c_2=10^{-5}$, $c_9=9.9$, $c_8=10^{-5}$ and scale is chosen to be $r^*=10\gg r_+$, (\textit{second row}) we change $c_9$ and assume that it is $0.9$, all other constants remain the same, (\textit{third row}) we change $c_8\to 10^{-2}$, all other constants remain the same}
    \label{fig:2}
\end{figure*}

As the second quantity of special interest, we consider heat capacity, which defines whether AdS black hole solution is valid or not. Numerical solutions for the heat capacity of our $f(Q)$-AdS black hole were placed in Figure (\ref{fig:2}). As expected, generally values of $C_P$ diverge at small event horizon radii $r_+$. However, generally with $r_+>r_s$, where $r_s$ is the singular point of $C_P$, heat capacity is positive which shows the physical behavior of our black hole solution. Moreover, as we noticed, with $c_9<0$, heat capacity values shrink. Similar behavior could be observed for the limit $\alpha\to0$. Judging by that, one could conclude that in quadratic $f(Q)$ gravity, AdS black holes have a bigger heat capacity than in the STEGR theory of gravity (GR analog). Finally, as $c_8\to\infty$, the logarithmic term in the metric potential starts to dominate nearby the event horizon, which leads to the non-trivial behavior of $C_P$. In that case, $r_s$ firstly shifts to the origin, then crosses some critical point $\alpha_c$ and starts shifting towards infinity, which restricts the physical plausibility of AdS BH solutions with small event horizon radii. It is worth noticing that $C_P$ numerical solutions are symmetrical with respect to $\alpha$, as well as the Hawking temperature investigated before.

\subsection{AdS BH phase transitions}
\begin{figure}[!htbp]
    \centering
    \includegraphics[width=\columnwidth]{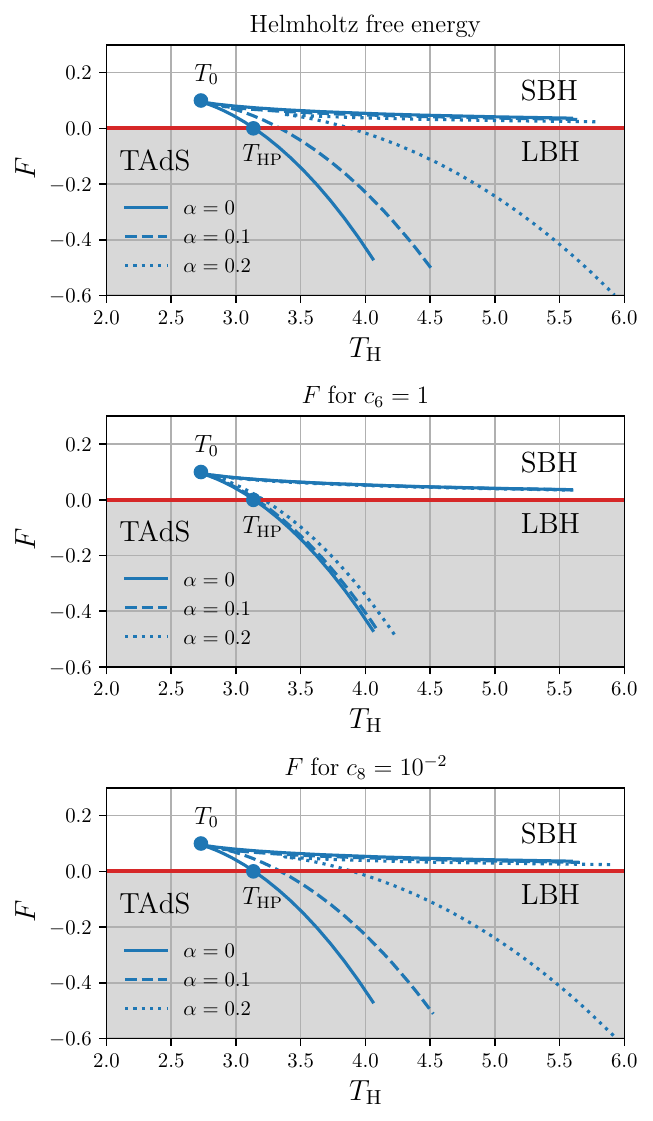}
    \caption{Helmholtz free energy for our AdS black hole solution with varying MOG free parameter $\alpha$ and varying constants of integration $c_8$, $c_9$. For the first plot, constants were fixed to $\Lambda=-3$, $c_2=10^{-5}$, $c_9=9.9$, $c_8=10^{-5}$. Moreover, on those plots we marked the point of Hawking-Page phase transition as $T_{\mathrm{HP}}$, small black hole solution as SBH and large as LBH, thermal AdS state as TAdS and finally Helmholtz free energy peak as $T_0$}
    \label{fig:3}
\end{figure}

This subsection is primarily devoted to the investigation of phase transitions of our $f(Q)$-AdS black hole. Previously mentioned and comprehensively studied heat capacity unveils the local behavior and stability of the black hole solution. In order to investigate our BH globally (phase transitions), one needs to use the so-called Helmholtz free energy, which is defined in terms of mass, Hawking temperature, and Bekenstein-Hawking entropy \cite{Sharif:2022ccc}:
\begin{equation}
    F = M - T_{\mathrm{H}}S.
\end{equation}
We plot the parametric plots within the phase plane $F-T_{\mathrm{H}}$ in Figure (\ref{fig:3}) for the usual choice of constants of integration and modified gravity free parameter $\alpha$, $\Lambda$-term. In the aforementioned figure, we marked the Large Black Hole solution (LBH) as the shaded area with $F<0$, Thermal AdS (TAdS) state as a line with $F=0$, peak Helmholtz free energy for Small Black Hole (SBH) as $T_0$ and Hawking-Page transition temperature (SBH$\to$TAdS$\to$LBH) as $T_{\mathrm{HP}}$. As we found out, $T_{\mathrm{HP}}$ shifts to infinity non-linearly with $|\alpha|\to\infty$. Moreover, with $c_9\to\infty$, change in Hawking-Page phase transition temperature slows down with growing $\alpha$, but peak temperature stays the same or gets smaller. As usual, more non-trivial behavior could be observed for big values of $c_8$ (so that the logarithmic term will dominate nearby event horizon). For such a case, $T_0$  will grow and $r_+$ at which our BH transits from SBH to LBH will also get smaller. Similar behavior of the Helmholtz free energy was observed for BPS (Bogomol'nyi-Prasad-Sommerfield) black hole embedded into the AdS$_4$ spacetime \cite{Ezroura:2021vrt} and for quintessential AdS black hole solution that arises from superstring theory \cite{Belhaj:2020pnh}, AdS-EGB (Einstein-Gauss-Bonnet) black holes \cite{Su:2019gby}.
\section{Exploring AdS black hole microstructures}\label{sec:4}
In this section, we investigate the microstructure of the AdS blackhole by introducing the Ruppeiner geometry. Here, we first provide a quick overview of the Ruppeiner geometry before calculating the charged AdS black hole's curvature scalar. Entropy S is used as the thermodynamical potential in Ruppeiner geometry and its variability $ dS^2$ is connected to the line element $ dl^2$ which represents the size of the separation between two nearby fluctuation states of the system of thermodynamics.

Let's think about an isolated thermodynamic system that is in equilibrium with total entropy S. Two subsystems make up the system: the little system we are interested in and its expansive surroundings which are denoted by $S_S$ and $S_E$ respectively, with the properties $S_S<<S_E\approx S$. So the system's overall entropy could be expressed as,
\begin{equation}
    S(x_0,x_1)=S_S(x_0,x_1)+S_E(x_0,x_1).
    \label{eq:35}
\end{equation}
In a system in equilibrium, the entropy achieves its peak locally. If we carry out the Taylor approximation in the vicinity of local maximum $x^\mu=x_0^\mu$, then equation (\ref{eq:35}) can be expressed as,
\begin{multline}
    S=S_0+\frac{\partial S_S}{\partial x^\mu} \biggr \rvert _{x^\mu_0} \Delta x^\mu _S + \frac{\partial S_E}{\partial x^\mu} \biggr \rvert _{x^\mu_0} \Delta x^\mu _E  \\ +\frac{1}{2} \frac{\partial^2 S_S}{\partial x^\mu \partial x^\nu} \biggr \rvert _{x^\mu_0} \Delta x^\mu _S  \Delta x^\nu _S+ \frac{1}{2} \frac{\partial^2 S_E}{\partial x^\mu \partial x^\nu} \biggr \rvert _{x^\mu_0} \Delta x^\mu _E  \Delta x^\nu _E +..... .
    \label{eq:36}
\end{multline}
Here $S_0$ is the zeroth order term that represents the local maximum entropy at the point $x^\mu _0$. Under the virtual change, the entropy of an isolated system in equilibrium remains constant.   
It is important to remember that $S_E$ is a thermodynamic significant quantity and is of the same magnitude as the system's overall entropy. As a result, it has a substantially lower derivative with regard to the intense quantity $x^\mu$ than $S_S$, which is thus not important.
As a result, $P(x_0,x_1)\propto e^{\frac{\Delta S}{K_B}}=e^{\frac{-1}{2}\Delta l^2}$  will be the probability of discovering the system in the intervals ($x_0$, $x_0 +dx_0$) and ($x_1$, $x_1 +dx_1$), where $K_B$ is the Boltzmann constant. From (\ref{eq:36}), the distance between two adjacent fluctuation states measured by the line element of Ruppeiner geometry can be expressed as,
 \cite{ruppeiner1995riemannian}
\begin{equation}
    d l^2=\frac{1}{k_B} g^{R}_{\mu \nu} \Delta x^{\mu} \Delta x^{\nu},
\end{equation}
where $k_B$ is the well-known Boltzmann constant, and the metric $g^{R}_{\mu\nu}$ is given by Ruppeiner(\cite{ruppeiner1995riemannian}) as,
\begin{equation}
    g^{R}_{\mu\nu}=-\frac{\partial^2 S(x)}{\partial x^{\mu} \partial x^{\nu}}.
\end{equation}
In which $S(x)$ is the entropy and $x^{\mu}$ are the extensive variables ascribed to a given thermodynamic system. For simplicity one typically starts with the definition of Weinhold's  metric \cite{weinhold1975metric} for evaluating the Ruppeiner metric as,
\begin{equation}
    g^{W}_{\mu\nu}=\frac{\partial^2 U(x)}{\partial x^{\mu} \partial x^{\nu}}.
\end{equation}
Where $U(x)$ is the internal energy of the BH system. Apart from that, it can be shown that the map
\begin{equation}
 dS^2_R=\frac{1}{T} dS^2_W 
\end{equation}
is used to show how the line elements in Weinhold and Ruppeiner geometry are linked conformally \cite{salamon1984relation,mrugala1984equivalence}.
Our goal is to investigate the microstructure interactions of the charged AdS black hole system using the tools of Ruppeiner geometry by treating the temperature $T$ and thermodynamic volume $V$ as the fluctuation variables. The metric $g^R_{\mu\nu}$ must be encrypted with the system's microscopic details because the line element  $ d l ^ 2$is related to the separation between two neighboring fluctuation states. By setting $k_B=1$, we could express the line element $ d l^2$ in ($S$,$P$) plane as, \cite{xu2020ruppeiner,ghosh2020thermodynamic}
\begin{equation}
     d l_R^2= \frac{1}{C_P} dS^2 +\frac{2}{T} \bigg(\frac{\partial T}{\partial P}\bigg)_{S} dS dP- \frac{V}{TB_S} dP^2 .
\end{equation}
Here $B_S$ represents adiabatic bulk modulus which can be expressed as $B_S= - V \bigg(\frac{\partial P}{\partial V}\bigg)_S$. In a similar way, the line element in ($T$,$V$) plane can be written as, \cite{xu2020ruppeiner,ghosh2020thermodynamic}
\begin{equation}
     d l_R^2= \frac{1}{T} \bigg(\frac{\partial P}{\partial V}\bigg)_T dV^2+\frac{2}{T} \bigg(\frac{\partial P}{\partial V} \bigg)_V dT dV +\frac{C_V}{T^2} dT^2,
\end{equation}
where $C_V=T \bigg(\frac{\partial S }{\partial T}\bigg)_V$ is the heat capacity under constant volume that can be found from the first law of thermodynamics for charged AdS black hole system. According to thermodynamic information geometry, the farther distant two thermodynamic states are from one another, the less likely it is that they will fluctuate. As a result, the line element $dl^2$ encapsulates data regarding the actual interaction of two tiny fluctuation states. The interaction information of the system's microstructure is generally embedded in the curvature scalar $R$ which we will calculate. The Riemannian geometry formulas can be used to calculate the Ricci scalar of geometry described by the Ruppeiner metric. The physical details of the microscopic interactions in a thermodynamic system are contained in this Ricci scalar, which we will refer to as the Ruppeiner curvature.  We may directly get the associated scalar curvature of the line element by applying the convention found in the literature (\cite{ruppeiner1995riemannian}) as

\begin{widetext}
\begin{align}
    \begin{aligned}
         R &=\frac{1}{2 C_V^2\left(\partial_V P\right)^2}\left\{T\left(\partial_V P\right)\left[\left(\partial_T C_V\right)\left(\partial_V P-T \partial_{T, V} P\right)+\left(\partial_V C_V\right)^2\right]\right. \\
        & \left.+C_V\left[\left(\partial_V P\right)^2+T\left(\left(\partial_V C_V\right)\left(\partial_V^2 P\right)-T\left(\partial_{T, V} P^2\right)\right)+2 T\left(\partial_V P\right)\left(T\left(\partial_{T, T, V} P\right)-\left(\partial_V^2 C_V\right)\right)\right]\right\}.
    \end{aligned}
\end{align}
\end{widetext}

The interactions which are repulsive or attractive can be classified by the positive sign of $R>0 $ or the negative sign of the $R<0$ respectively. If the Ruppeiner curvature disappears, there probably isn't any real interaction going on between the tiny molecules i.e. the interaction of attraction and repulsion achieves its equilibrium. Additionally, it is hypothesized that the Ruppeiner scalar diverges at the critical point.
\begin{figure}[!htbp]
    \centering
    \includegraphics[width=\columnwidth]{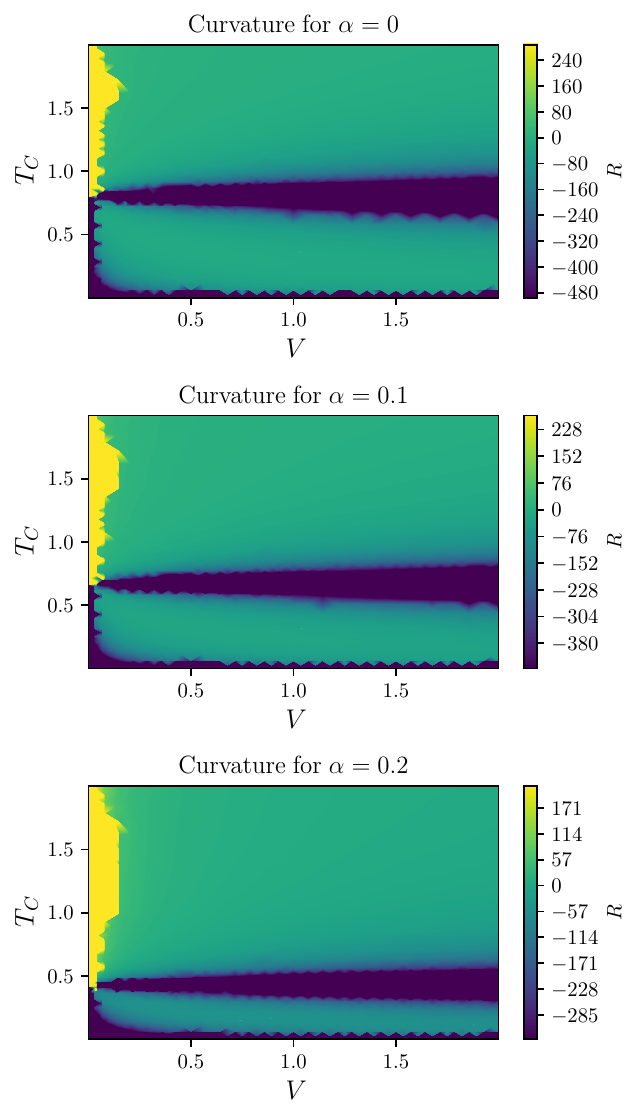}
    \caption{Ruppeiner curvature for AdS-$f(Q)$ wormhole with $\Lambda=-3$, $c_2=10^{-5}$, $c_9=9.9$, $c_8=10^{-5}$ and varying values of constant Hawking temperature, $\alpha$ and BH volume}
    \label{fig:last}
\end{figure}

We carefully plot the numerical solutions for Ruppeiner curvature scalar $R$ in Figure (\ref{fig:last}) for our modified AdS BH solution with varying values of constant Hawking temperature, volume, and $\alpha$. It could be noticed for all cases that we considered, from some point $R|_{T_C=T_p}$, interactions are only repulsive or for some particular cases, even the complete absence of such interactions could be observed for higher values of volume $V$. On the other hand, for $T_{C}<T_p$, singular points are present, and nearby those points interactions have a strongly repulsive manner. It is interesting that $T_p$ shifts towards origin for $\alpha\to\infty$.
\section{Conclusions}\label{sec:5}
In this paper, we presented the comprehensive study of the thermodynamical behavior of the AdS black hole in the symmetric teleparallel theory of gravitation. We investigated extensively various thermodynamical quantities as well as black hole microstructure. 

As the first quantity of special interest in our paper, we have chosen the Hawking temperature - the temperature of Hawking radiation measured by a stationary observer at the spatial infinity. Such radiation cannot classically occur within the BH spacetime, and therefore it is formed due to the presence of some quantum effects. We numerically derived the Hawking temperature for our black hole solution and showed those solutions graphically in Figure (\ref{fig:1}). It was found that values of Hawking temperature do not change with the variation of the constant of integration $c_9$ but do get bigger with growing values of both $\alpha$ and $c_8$ (for $c_8\gg0$, log term in the metric tensor will dominate nearby event horizon). The obvious conclusion is that our BH solution will evaporate faster if a quadratic term in the STEGR action will dominate over the linear one.

The next important thermodynamical quantity that was probed is the well-known heat capacity at the constant thermodynamical pressure $P$. As expected, at small values of event horizon radii $r_+$, heat capacity diverges. But, if $r_+>r_s$ (where $r_s$ is the point at which heat-capacity diverges), heat-capacity will be generally positive, which could result in a physically viable behavior of the black hole model. Besides, for $c_9<0\land \alpha \to 0$, $C_P\to0$. However, if $c_8$ is sufficiently big so that the log term will dominate the nearby event horizon, the behavior of the heat capacity starts to become highly non-trivial. With $c_8\to\infty$, the singular point of the heat capacity firstly shifts up to the origin, then to positive infinity. Such behavior constrains the horizon radii to be small. For a graphical representation of the aforementioned results, refer to Figure (\ref{fig:2}). 

Apart from the Hawking temperature and heat capacity, we as well properly investigated the behavior of free Helmholtz energy. As we unveiled, the Hawking-Page phase transition point shifts up to the positive infinity in the case of $|\alpha|\to\infty$. On the other hand, if $c_9\to\infty$, $T_{\mathrm{HP}}$ starts to shift towards $+\infty$ slower with the growing trend of $\alpha$ and peak temperature $T_0$ stays the same or gets smaller. Parametric plots for Helmholtz free energy versus Hawking temperature are located in Figure (\ref{fig:3}) respectively.

Finally, the black hole microstructures were studied by the construction of metric tensors with $P$ and $T_{\mathrm{H}}$ as coordinates. Afterward, we derived the Ricci scalar curvature (Ruppeiner curvature) from the resulting metric tensor and carefully plotted the results in Figure (\ref{fig:last}). From such a graphical representation we could conclude that above some point $T_p$, interactions are strongly repulsive or vanish. However, for the case with Hawking temperature being smaller than $T_p$, we have singular points and strongly attractive regions. Besides, it could be easily noticed that $T_p\to0$ as $\alpha\to\infty$.

Now, it will be a handful to compare our results with the ones, that are already present in the literature. Firstly, it is worth noticing that the behavior of Hawking temperature for the black hole model of our consideration is very similar to such cases as  Kerr-Newman-AdS, AdS-EGB, non-minimal derivative coupling black holes (see \cite{Caldarelli:1999xj,Cvetic:2001bk,Rinaldi:2012vy}). On the other hand, in the case of heat capacity, our results are very similar to such BH solutions as quantum-corrected Schild-AdS black holes, GUP-corrected black holes with topological defects, and rotating Lifshitz-like black holes within $f(R)$ modified theory of gravitation \cite{Chen:2023pcv,Karmakar:2023mhs,Jafarzade:2023yof}. Each of those aforementioned solutions had heat capacity, that diverges at small $r_+$ and that asymptotically converges at some constant value. Finally, one can also compare the behavior of Helmholtz's free energy. In that situation, we have found that Renyi-entropy black hole, rotating black hole/ring, and Gauss-Bonnet black hole solutions \cite{Czinner:2015eyk,Altamirano:2014tva,Su:2019gby} represent the behavior, very similar to our case in sense of $F$. Therefore, one can conclude that our modified gravity black hole solution is not unique and rather represents a large set of various alike solutions with various cosmological backgrounds. According to the accepted theory, black holes are almost the simplest structures in general relativity. However, with the development of black hole thermodynamics, we have begun to understand them as extremely complicated thermal systems that are rarely in equilibrium and have an incredibly large number of internal states. In our study on thermodynamical quantities of BH, we can conclude that in quadratic $f(Q)$ gravity, AdS BH has a bigger heat capacity than in the STEGR theory of gravity (GR analog). Apart from that, from the study of BH microstructure, we found in the graphical representation that above a specific point of $T_P$, the interaction between the particles is strongly repulsive, which shows almost the same behavior as the microstructures of Euler-Heisenberg BH in work \cite{yu2023thermodynamics}

In our future studies, it would be interesting to obtain and investigate AdS BH solutions within other $f(Q)$ gravity theories and in the presence of additional matter fields, such as scalar field, spin 1/2 Dirac spinors, spin-2 massive fields, etc. Also, it would be of special interest to test the point-particle dynamics around such black hole solutions and derive effective potential.

\section*{Data Availability Statement}
There are no new data associated with this article.

\section*{Acknowledgement}
Sokolik O. performed the work in the frame of the ``Mathematical modeling in interdisciplinary research of processes and systems based on intelligent supercomputer, grid, and cloudS technologies" program of the NAS of Ukraine. PKS acknowledges the National Board for Higher Mathematics (NBHM) under the Department of Atomic Energy (DAE), Govt. of India, for financial support to carry out the Research project No.: 02011/3/2022 NBHM(R.P.)/R \& D II/2152 Dt.14.02.2022. We are very grateful to the honorable referee and the editor for the illuminating suggestions that have significantly improved our research quality and presentation.

\end{CJK*}

\begin{thebibliography}{53}%
\makeatletter
\providecommand \@ifxundefined [1]{%
 \@ifx{#1\undefined}
}%
\providecommand \@ifnum [1]{%
 \ifnum #1\expandafter \@firstoftwo
 \else \expandafter \@secondoftwo
 \fi
}%
\providecommand \@ifx [1]{%
 \ifx #1\expandafter \@firstoftwo
 \else \expandafter \@secondoftwo
 \fi
}%
\providecommand \natexlab [1]{#1}%
\providecommand \enquote  [1]{``#1''}%
\providecommand \bibnamefont  [1]{#1}%
\providecommand \bibfnamefont [1]{#1}%
\providecommand \citenamefont [1]{#1}%
\providecommand \href@noop [0]{\@secondoftwo}%
\providecommand \href [0]{\begingroup \@sanitize@url \@href}%
\providecommand \@href[1]{\@@startlink{#1}\@@href}%
\providecommand \@@href[1]{\endgroup#1\@@endlink}%
\providecommand \@sanitize@url [0]{\catcode `\\12\catcode `\$12\catcode
  `\&12\catcode `\#12\catcode `\^12\catcode `\_12\catcode `\%12\relax}%
\providecommand \@@startlink[1]{}%
\providecommand \@@endlink[0]{}%
\providecommand \url  [0]{\begingroup\@sanitize@url \@url }%
\providecommand \@url [1]{\endgroup\@href {#1}{\urlprefix }}%
\providecommand \urlprefix  [0]{URL }%
\providecommand \Eprint [0]{\href }%
\providecommand \doibase [0]{https://doi.org/}%
\providecommand \selectlanguage [0]{\@gobble}%
\providecommand \bibinfo  [0]{\@secondoftwo}%
\providecommand \bibfield  [0]{\@secondoftwo}%
\providecommand \translation [1]{[#1]}%
\providecommand \BibitemOpen [0]{}%
\providecommand \bibitemStop [0]{}%
\providecommand \bibitemNoStop [0]{.\EOS\space}%
\providecommand \EOS [0]{\spacefactor3000\relax}%
\providecommand \BibitemShut  [1]{\csname bibitem#1\endcsname}%
\let\auto@bib@innerbib\@empty
\bibitem [{\citenamefont {Bardeen}\ \emph {et~al.}(1973)\citenamefont
  {Bardeen}, \citenamefont {Carter},\ and\ \citenamefont
  {Hawking}}]{bardeen1973four}%
  \BibitemOpen
  \bibfield  {author} {\bibinfo {author} {\bibfnamefont {J.~M.}\ \bibnamefont
  {Bardeen}}, \bibinfo {author} {\bibfnamefont {B.}~\bibnamefont {Carter}},\
  and\ \bibinfo {author} {\bibfnamefont {S.~W.}\ \bibnamefont {Hawking}},\
  }\bibfield  {title} {\bibinfo {title} {The four laws of black hole
  mechanics},\ }\href {https://doi.org/http://dx.doi.org/10.1007/BF01645742}
  {\bibfield  {journal} {\bibinfo  {journal} {Communications in mathematical
  physics}\ }\textbf {\bibinfo {volume} {31}},\ \bibinfo {pages} {161}
  (\bibinfo {year} {1973})}\BibitemShut {NoStop}%
\bibitem [{\citenamefont {Hawking}(1975)}]{hawking1975particle}%
  \BibitemOpen
  \bibfield  {author} {\bibinfo {author} {\bibfnamefont {S.~W.}\ \bibnamefont
  {Hawking}},\ }\bibfield  {title} {\bibinfo {title} {Particle creation by
  black holes},\ }in\ \href@noop {} {\emph {\bibinfo {booktitle} {Euclidean
  quantum gravity}}}\ (\bibinfo  {publisher} {World Scientific},\ \bibinfo
  {year} {1975})\ pp.\ \bibinfo {pages} {167--188}\BibitemShut {NoStop}%
\bibitem [{\citenamefont {Hawking}\ and\ \citenamefont
  {Page}(1983)}]{hawking1983thermodynamics}%
  \BibitemOpen
  \bibfield  {author} {\bibinfo {author} {\bibfnamefont {S.~W.}\ \bibnamefont
  {Hawking}}\ and\ \bibinfo {author} {\bibfnamefont {D.~N.}\ \bibnamefont
  {Page}},\ }\bibfield  {title} {\bibinfo {title} {Thermodynamics of black
  holes in anti-de sitter space},\ }\href@noop {} {\bibfield  {journal}
  {\bibinfo  {journal} {Communications in Mathematical Physics}\ }\textbf
  {\bibinfo {volume} {87}},\ \bibinfo {pages} {577} (\bibinfo {year}
  {1983})}\BibitemShut {NoStop}%
\bibitem [{\citenamefont {Eom}\ \emph {et~al.}(2022)\citenamefont {Eom},
  \citenamefont {Jung},\ and\ \citenamefont {Kim}}]{eom2022hawking}%
  \BibitemOpen
  \bibfield  {author} {\bibinfo {author} {\bibfnamefont {H.}~\bibnamefont
  {Eom}}, \bibinfo {author} {\bibfnamefont {S.}~\bibnamefont {Jung}},\ and\
  \bibinfo {author} {\bibfnamefont {W.}~\bibnamefont {Kim}},\ }\bibfield
  {title} {\bibinfo {title} {Hawking-page phase transition of the schwarzschild
  ads black hole with the effective tolman temperature},\ }\href
  {arXiv:2205.09938v2} {\bibfield  {journal} {\bibinfo  {journal} {Journal of
  Cosmology and Astroparticle Physics}\ }\textbf {\bibinfo {volume}
  {2022}}\bibinfo  {number} { (09)},\ \bibinfo {pages} {053}}\BibitemShut
  {NoStop}%
\bibitem [{\citenamefont {Pourhassan}\ \emph
  {et~al.}(2020{\natexlab{a}})\citenamefont {Pourhassan}, \citenamefont
  {{\"O}vg{\"u}n},\ and\ \citenamefont {Sakall{\i}}}]{pourhassan2020pv}%
  \BibitemOpen
\bibfield  {number} {  }\bibfield  {author} {\bibinfo {author} {\bibfnamefont
  {B.}~\bibnamefont {Pourhassan}}, \bibinfo {author} {\bibfnamefont
  {A.}~\bibnamefont {{\"O}vg{\"u}n}},\ and\ \bibinfo {author} {\bibfnamefont
  {{\.I}.}~\bibnamefont {Sakall{\i}}},\ }\bibfield  {title} {\bibinfo {title}
  {Pv criticality of achucarro--ortiz black hole in the presence of
  higher-order quantum and gup corrections},\ }\href
  {https://doi.org/https://doi.org/10.1142/S021988782050156X} {\bibfield
  {journal} {\bibinfo  {journal} {International Journal of Geometric Methods in
  Modern Physics}\ }\textbf {\bibinfo {volume} {17}},\ \bibinfo {pages}
  {2050156} (\bibinfo {year} {2020}{\natexlab{a}})}\BibitemShut {NoStop}%
\bibitem [{\citenamefont {Ghosh}\ \emph {et~al.}(2020)\citenamefont {Ghosh},
  \citenamefont {Kumar},\ and\ \citenamefont {Singh}}]{Ghosh:2020tgy}%
  \BibitemOpen
  \bibfield  {author} {\bibinfo {author} {\bibfnamefont {S.~G.}\ \bibnamefont
  {Ghosh}}, \bibinfo {author} {\bibfnamefont {A.}~\bibnamefont {Kumar}},\ and\
  \bibinfo {author} {\bibfnamefont {D.~V.}\ \bibnamefont {Singh}},\ }\bibfield
  {title} {\bibinfo {title} {{Anti-de Sitter Hayward black holes in
  Einstein\textendash{}Gauss\textendash{}Bonnet gravity}},\ }\href
  {https://doi.org/10.1016/j.dark.2020.100660} {\bibfield  {journal} {\bibinfo
  {journal} {Phys. Dark Univ.}\ }\textbf {\bibinfo {volume} {30}},\ \bibinfo
  {pages} {100660} (\bibinfo {year} {2020})}\BibitemShut {NoStop}%
\bibitem [{\citenamefont {Javed}\ \emph {et~al.}(2023)\citenamefont {Javed},
  \citenamefont {Atique}, \citenamefont {Pantig},\ and\ \citenamefont
  {{\"O}vg{\"u}n}}]{javed2023weak}%
  \BibitemOpen
  \bibfield  {author} {\bibinfo {author} {\bibfnamefont {W.}~\bibnamefont
  {Javed}}, \bibinfo {author} {\bibfnamefont {M.}~\bibnamefont {Atique}},
  \bibinfo {author} {\bibfnamefont {R.~C.}\ \bibnamefont {Pantig}},\ and\
  \bibinfo {author} {\bibfnamefont {A.}~\bibnamefont {{\"O}vg{\"u}n}},\
  }\bibfield  {title} {\bibinfo {title} {Weak lensing, hawking radiation and
  greybody factor bound by a charged black holes with non-linear
  electrodynamics corrections},\ }\href
  {https://doi.org/https://doi.org/10.1142/S0219887823500408} {\bibfield
  {journal} {\bibinfo  {journal} {International Journal of Geometric Methods in
  Modern Physics}\ }\textbf {\bibinfo {volume} {20}},\ \bibinfo {pages}
  {2350040} (\bibinfo {year} {2023})}\BibitemShut {NoStop}%
\bibitem [{\citenamefont {Sood}\ \emph {et~al.}()\citenamefont {Sood},
  \citenamefont {Kumar}, \citenamefont {Singh},\ and\ \citenamefont
  {Ghosh}}]{sood2022thermodynamic}%
  \BibitemOpen
  \bibfield  {author} {\bibinfo {author} {\bibfnamefont {A.}~\bibnamefont
  {Sood}}, \bibinfo {author} {\bibfnamefont {A.}~\bibnamefont {Kumar}},
  \bibinfo {author} {\bibfnamefont {J.}~\bibnamefont {Singh}},\ and\ \bibinfo
  {author} {\bibfnamefont {S.~G.}\ \bibnamefont {Ghosh}},\ }\bibfield  {title}
  {\bibinfo {title} {Thermodynamic stability and p--v criticality of
  nonsingular-ads black holes endowed with clouds of strings},\ }\bibfield
  {journal} {\bibinfo  {journal} {The European Physical Journal C}\ }\textbf
  {\bibinfo {volume} {82}},\ \href
  {https://doi.org/https://doi.org/10.1140/epjc/s10052-022-10181-8}
  {https://doi.org/10.1140/epjc/s10052-022-10181-8}\BibitemShut {NoStop}%
\bibitem [{\citenamefont {\"Ovg\"un}(2018)}]{Ovgun:2017bgx}%
  \BibitemOpen
  \bibfield  {author} {\bibinfo {author} {\bibfnamefont {A.}~\bibnamefont
  {\"Ovg\"un}},\ }\bibfield  {title} {\bibinfo {title} {{$P-v$ criticality of a
  specific black hole in $f(R)$ gravity coupled with Yang-Mills field}},\
  }\href {https://doi.org/10.1155/2018/8153721} {\bibfield  {journal} {\bibinfo
   {journal} {Adv. High Energy Phys.}\ }\textbf {\bibinfo {volume} {2018}},\
  \bibinfo {pages} {8153721} (\bibinfo {year} {2018})},\ \Eprint
  {https://arxiv.org/abs/1710.06795} {arXiv:1710.06795 [gr-qc]} \BibitemShut
  {NoStop}%
\bibitem [{\citenamefont {Pourhassan}\ \emph
  {et~al.}(2020{\natexlab{b}})\citenamefont {Pourhassan}, \citenamefont
  {\"Ovg\"un},\ and\ \citenamefont {Sakall\i{}}}]{Pourhassan:2018chj}%
  \BibitemOpen
  \bibfield  {author} {\bibinfo {author} {\bibfnamefont {B.}~\bibnamefont
  {Pourhassan}}, \bibinfo {author} {\bibfnamefont {A.}~\bibnamefont
  {\"Ovg\"un}},\ and\ \bibinfo {author} {\bibfnamefont {I.}~\bibnamefont
  {Sakall\i{}}},\ }\bibfield  {title} {\bibinfo {title} {{PV criticality of
  Achucarro-Ortiz black hole in the presence of higher order quantum and GUP
  corrections}},\ }\href {https://doi.org/10.1142/S021988782050156X} {\bibfield
   {journal} {\bibinfo  {journal} {Int. J. Geom. Meth. Mod. Phys.}\ }\textbf
  {\bibinfo {volume} {17}},\ \bibinfo {pages} {2050156} (\bibinfo {year}
  {2020}{\natexlab{b}})},\ \Eprint {https://arxiv.org/abs/1811.02193}
  {arXiv:1811.02193 [gr-qc]} \BibitemShut {NoStop}%
\bibitem [{\citenamefont {Kastor}\ \emph {et~al.}()\citenamefont {Kastor},
  \citenamefont {Ray},\ and\ \citenamefont {Traschen}}]{kastor2009enthalpy}%
  \BibitemOpen
  \bibfield  {author} {\bibinfo {author} {\bibfnamefont {D.}~\bibnamefont
  {Kastor}}, \bibinfo {author} {\bibfnamefont {S.}~\bibnamefont {Ray}},\ and\
  \bibinfo {author} {\bibfnamefont {J.}~\bibnamefont {Traschen}},\ }\bibfield
  {title} {\bibinfo {title} {Enthalpy and the mechanics of ads black holes},\
  }\bibfield  {journal} {\bibinfo  {journal} {Classical and Quantum Gravity}\
  }\textbf {\bibinfo {volume} {26}},\ \href
  {https://doi.org/https://doi.org/10.48550/arXiv.0904.2765}
  {https://doi.org/10.48550/arXiv.0904.2765}\BibitemShut {NoStop}%
\bibitem [{\citenamefont {Dolan}()}]{dolan2011pressure}%
  \BibitemOpen
  \bibfield  {author} {\bibinfo {author} {\bibfnamefont {B.~P.}\ \bibnamefont
  {Dolan}},\ }\bibfield  {title} {\bibinfo {title} {Pressure and volume in the
  first law of black hole thermodynamics},\ }\bibfield  {journal} {\bibinfo
  {journal} {Classical and Quantum Gravity}\ }\href
  {https://doi.org/arXiv:1106.6260} {arXiv:1106.6260}\BibitemShut {NoStop}%
\bibitem [{\citenamefont {Kubiz{\v{n}}{\'a}k}\ and\ \citenamefont
  {Mann}(2012)}]{kubizvnak2012p}%
  \BibitemOpen
  \bibfield  {author} {\bibinfo {author} {\bibfnamefont {D.}~\bibnamefont
  {Kubiz{\v{n}}{\'a}k}}\ and\ \bibinfo {author} {\bibfnamefont {R.~B.}\
  \bibnamefont {Mann}},\ }\bibfield  {title} {\bibinfo {title} {P- v
  criticality of charged ads black holes},\ }\href@noop {} {\bibfield
  {journal} {\bibinfo  {journal} {Journal of High Energy Physics}\ }\textbf
  {\bibinfo {volume} {2012}},\ \bibinfo {pages} {1} (\bibinfo {year}
  {2012})}\BibitemShut {NoStop}%
\bibitem [{\citenamefont {Cai}\ \emph {et~al.}()\citenamefont {Cai},
  \citenamefont {Cao}, \citenamefont {Li},\ and\ \citenamefont
  {Yang}}]{cai2013pv}%
  \BibitemOpen
  \bibfield  {author} {\bibinfo {author} {\bibfnamefont {R.-G.}\ \bibnamefont
  {Cai}}, \bibinfo {author} {\bibfnamefont {L.-M.}\ \bibnamefont {Cao}},
  \bibinfo {author} {\bibfnamefont {L.}~\bibnamefont {Li}},\ and\ \bibinfo
  {author} {\bibfnamefont {R.-Q.}\ \bibnamefont {Yang}},\ }\bibfield  {title}
  {\bibinfo {title} {Pv criticality in the extended phase space of gauss-bonnet
  black holes in ads space},\ }\bibfield  {journal} {\bibinfo  {journal}
  {Journal of High Energy Physics}\ }\href
  {https://doi.org/https://doi.org/10.1007/JHEP09(2013)005}
  {https://doi.org/10.1007/JHEP09(2013)005}\BibitemShut {NoStop}%
\bibitem [{\citenamefont {Chamblin}\ \emph {et~al.}(1999)\citenamefont
  {Chamblin}, \citenamefont {Emparan}, \citenamefont {Johnson},\ and\
  \citenamefont {Myers}}]{chamblin1999charged}%
  \BibitemOpen
  \bibfield  {author} {\bibinfo {author} {\bibfnamefont {A.}~\bibnamefont
  {Chamblin}}, \bibinfo {author} {\bibfnamefont {R.}~\bibnamefont {Emparan}},
  \bibinfo {author} {\bibfnamefont {C.~V.}\ \bibnamefont {Johnson}},\ and\
  \bibinfo {author} {\bibfnamefont {R.~C.}\ \bibnamefont {Myers}},\ }\bibfield
  {title} {\bibinfo {title} {Charged ads black holes and catastrophic
  holography},\ }\href {https://doi.org/arXiv:hep-th/9902170v2} {\bibfield
  {journal} {\bibinfo  {journal} {Physical Review D}\ }\textbf {\bibinfo
  {volume} {60}},\ \bibinfo {pages} {064018} (\bibinfo {year}
  {1999})}\BibitemShut {NoStop}%
\bibitem [{\citenamefont {Bekenstein}(1973)}]{B1}%
  \BibitemOpen
  \bibfield  {author} {\bibinfo {author} {\bibfnamefont {J.~D.}\ \bibnamefont
  {Bekenstein}},\ }\bibfield  {title} {\bibinfo {title} {Black holes and
  entropy},\ }\href@noop {} {\bibfield  {journal} {\bibinfo  {journal}
  {Physical Review D}\ }\textbf {\bibinfo {volume} {7}},\ \bibinfo {pages}
  {2333} (\bibinfo {year} {1973})}\BibitemShut {NoStop}%
\bibitem [{\citenamefont {Bekenstein}(1974)}]{B2}%
  \BibitemOpen
  \bibfield  {author} {\bibinfo {author} {\bibfnamefont {J.~D.}\ \bibnamefont
  {Bekenstein}},\ }\bibfield  {title} {\bibinfo {title} {Generalized second law
  of thermodynamics in black-hole physics},\ }\href@noop {} {\bibfield
  {journal} {\bibinfo  {journal} {Physical Review D}\ }\textbf {\bibinfo
  {volume} {9}},\ \bibinfo {pages} {3292} (\bibinfo {year} {1974})}\BibitemShut
  {NoStop}%
\bibitem [{\citenamefont {Shen}\ \emph {et~al.}()\citenamefont {Shen},
  \citenamefont {Cai}, \citenamefont {Wang},\ and\ \citenamefont
  {Su}}]{shen2007thermodynamic}%
  \BibitemOpen
  \bibfield  {author} {\bibinfo {author} {\bibfnamefont {J.}~\bibnamefont
  {Shen}}, \bibinfo {author} {\bibfnamefont {R.-G.}\ \bibnamefont {Cai}},
  \bibinfo {author} {\bibfnamefont {B.}~\bibnamefont {Wang}},\ and\ \bibinfo
  {author} {\bibfnamefont {R.-K.}\ \bibnamefont {Su}},\ }\bibfield  {title}
  {\bibinfo {title} {Thermodynamic geometry and critical behavior of black
  holes},\ }\bibfield  {journal} {\bibinfo  {journal} {International Journal of
  Modern Physics A}\ }\textbf {\bibinfo {volume} {22}},\ \href
  {https://doi.org/arXiv:gr-qc/0512035} {arXiv:gr-qc/0512035}\BibitemShut
  {NoStop}%
\bibitem [{\citenamefont {Alfaia}\ \emph {et~al.}(2022)\citenamefont {Alfaia},
  \citenamefont {Lobo},\ and\ \citenamefont {Brito}}]{alfaia2022central}%
  \BibitemOpen
  \bibfield  {author} {\bibinfo {author} {\bibfnamefont {R.}~\bibnamefont
  {Alfaia}}, \bibinfo {author} {\bibfnamefont {I.}~\bibnamefont {Lobo}},\ and\
  \bibinfo {author} {\bibfnamefont {L.}~\bibnamefont {Brito}},\ }\bibfield
  {title} {\bibinfo {title} {Central charge criticality of charged ads black
  hole surrounded by different fluids},\ }\href@noop {} {\bibfield  {journal}
  {\bibinfo  {journal} {The European Physical Journal Plus}\ }\textbf {\bibinfo
  {volume} {137}},\ \bibinfo {pages} {402} (\bibinfo {year}
  {2022})}\BibitemShut {NoStop}%
\bibitem [{\citenamefont {Wei}\ and\ \citenamefont
  {Liu}(2022)}]{wei2022microstructure}%
  \BibitemOpen
  \bibfield  {author} {\bibinfo {author} {\bibfnamefont {S.-W.}\ \bibnamefont
  {Wei}}\ and\ \bibinfo {author} {\bibfnamefont {Y.-X.}\ \bibnamefont {Liu}},\
  }\bibfield  {title} {\bibinfo {title} {The microstructure and ruppeiner
  geometry of charged anti-de sitter black holes in gauss--bonnet gravity: from
  the critical point to the triple point},\ }\href@noop {} {\bibfield
  {journal} {\bibinfo  {journal} {Communications in Theoretical Physics}\
  }\textbf {\bibinfo {volume} {74}},\ \bibinfo {pages} {095402} (\bibinfo
  {year} {2022})}\BibitemShut {NoStop}%
\bibitem [{\citenamefont {Wei}\ \emph {et~al.}(2019)\citenamefont {Wei},
  \citenamefont {Liu},\ and\ \citenamefont {Mann}}]{wei2019ruppeiner}%
  \BibitemOpen
  \bibfield  {author} {\bibinfo {author} {\bibfnamefont {S.-W.}\ \bibnamefont
  {Wei}}, \bibinfo {author} {\bibfnamefont {Y.-X.}\ \bibnamefont {Liu}},\ and\
  \bibinfo {author} {\bibfnamefont {R.~B.}\ \bibnamefont {Mann}},\ }\bibfield
  {title} {\bibinfo {title} {Ruppeiner geometry, phase transitions, and the
  microstructure of charged ads black holes},\ }\href@noop {} {\bibfield
  {journal} {\bibinfo  {journal} {Physical Review D}\ }\textbf {\bibinfo
  {volume} {100}},\ \bibinfo {pages} {124033} (\bibinfo {year}
  {2019})}\BibitemShut {NoStop}%
\bibitem [{\citenamefont {Nashed}(2021)}]{nashed2021new}%
  \BibitemOpen
  \bibfield  {author} {\bibinfo {author} {\bibfnamefont {G.}~\bibnamefont
  {Nashed}},\ }\bibfield  {title} {\bibinfo {title} {New rotating ads/ds black
  holes in f (r) gravity},\ }\href@noop {} {\bibfield  {journal} {\bibinfo
  {journal} {Physics Letters B}\ }\textbf {\bibinfo {volume} {815}},\ \bibinfo
  {pages} {136133} (\bibinfo {year} {2021})}\BibitemShut {NoStop}%
\bibitem [{\citenamefont {Li}\ \emph {et~al.}(2018)\citenamefont {Li},
  \citenamefont {He}, \citenamefont {Ding}, \citenamefont {Hu},\ and\
  \citenamefont {Deng}}]{li2018thermodynamics}%
  \BibitemOpen
  \bibfield  {author} {\bibinfo {author} {\bibfnamefont {P.}~\bibnamefont
  {Li}}, \bibinfo {author} {\bibfnamefont {M.}~\bibnamefont {He}}, \bibinfo
  {author} {\bibfnamefont {J.-C.}\ \bibnamefont {Ding}}, \bibinfo {author}
  {\bibfnamefont {X.-R.}\ \bibnamefont {Hu}},\ and\ \bibinfo {author}
  {\bibfnamefont {J.-B.}\ \bibnamefont {Deng}},\ }\bibfield  {title} {\bibinfo
  {title} {Thermodynamics of charged ads black holes in rainbow gravity},\
  }\href {https://doi.org/https://doi.org/10.1155/2018/1043639} {\bibfield
  {journal} {\bibinfo  {journal} {Advances in High Energy Physics}\ }\textbf
  {\bibinfo {volume} {2018}},\ \bibinfo {pages} {1} (\bibinfo {year}
  {2018})}\BibitemShut {NoStop}%
\bibitem [{\citenamefont {Heisenberg}\ and\ \citenamefont
  {Tsujikawa}(2018)}]{heisenberg2018hairy}%
  \BibitemOpen
  \bibfield  {author} {\bibinfo {author} {\bibfnamefont {L.}~\bibnamefont
  {Heisenberg}}\ and\ \bibinfo {author} {\bibfnamefont {S.}~\bibnamefont
  {Tsujikawa}},\ }\bibfield  {title} {\bibinfo {title} {Hairy black hole
  solutions in u (1) gauge-invariant scalar--vector--tensor theories},\
  }\href@noop {} {\bibfield  {journal} {\bibinfo  {journal} {Physics Letters
  B}\ }\textbf {\bibinfo {volume} {780}},\ \bibinfo {pages} {638} (\bibinfo
  {year} {2018})}\BibitemShut {NoStop}%
\bibitem [{\citenamefont {Heisenberg}\ \emph {et~al.}(2017)\citenamefont
  {Heisenberg}, \citenamefont {Kase}, \citenamefont {Minamitsuji},\ and\
  \citenamefont {Tsujikawa}}]{heisenberg2017hairy}%
  \BibitemOpen
  \bibfield  {author} {\bibinfo {author} {\bibfnamefont {L.}~\bibnamefont
  {Heisenberg}}, \bibinfo {author} {\bibfnamefont {R.}~\bibnamefont {Kase}},
  \bibinfo {author} {\bibfnamefont {M.}~\bibnamefont {Minamitsuji}},\ and\
  \bibinfo {author} {\bibfnamefont {S.}~\bibnamefont {Tsujikawa}},\ }\bibfield
  {title} {\bibinfo {title} {Hairy black-hole solutions in generalized proca
  theories},\ }\href@noop {} {\bibfield  {journal} {\bibinfo  {journal}
  {Physical Review D}\ }\textbf {\bibinfo {volume} {96}},\ \bibinfo {pages}
  {084049} (\bibinfo {year} {2017})}\BibitemShut {NoStop}%
\bibitem [{\citenamefont {D’Ambrosio}\ \emph {et~al.}(2022)\citenamefont
  {D’Ambrosio}, \citenamefont {Fell}, \citenamefont {Heisenberg},\ and\
  \citenamefont {Kuhn}}]{d2022black}%
  \BibitemOpen
  \bibfield  {author} {\bibinfo {author} {\bibfnamefont {F.}~\bibnamefont
  {D’Ambrosio}}, \bibinfo {author} {\bibfnamefont {S.~D.}\ \bibnamefont
  {Fell}}, \bibinfo {author} {\bibfnamefont {L.}~\bibnamefont {Heisenberg}},\
  and\ \bibinfo {author} {\bibfnamefont {S.}~\bibnamefont {Kuhn}},\ }\bibfield
  {title} {\bibinfo {title} {Black holes in f (q) gravity},\ }\href@noop {}
  {\bibfield  {journal} {\bibinfo  {journal} {Physical Review D}\ }\textbf
  {\bibinfo {volume} {105}},\ \bibinfo {pages} {024042} (\bibinfo {year}
  {2022})}\BibitemShut {NoStop}%
\bibitem [{\citenamefont {Ortín}(2015)}]{ortin2015}%
  \BibitemOpen
  \bibfield  {author} {\bibinfo {author} {\bibfnamefont {T.}~\bibnamefont
  {Ortín}},\ }\href {https://doi.org/10.1017/CBO9781139019750} {\emph
  {\bibinfo {title} {Gravity and Strings}}},\ \bibinfo {edition} {2nd}\ ed.,\
  Cambridge Monographs on Mathematical Physics\ (\bibinfo  {publisher}
  {Cambridge University Press},\ \bibinfo {year} {2015})\BibitemShut {NoStop}%
\bibitem [{\citenamefont {Xu}\ \emph {et~al.}({\natexlab{a}})\citenamefont
  {Xu}, \citenamefont {Li}, \citenamefont {Harko},\ and\ \citenamefont
  {Liang}}]{Xu2019}%
  \BibitemOpen
  \bibfield  {author} {\bibinfo {author} {\bibfnamefont {Y.}~\bibnamefont
  {Xu}}, \bibinfo {author} {\bibfnamefont {G.}~\bibnamefont {Li}}, \bibinfo
  {author} {\bibfnamefont {T.}~\bibnamefont {Harko}},\ and\ \bibinfo {author}
  {\bibfnamefont {S.-D.}\ \bibnamefont {Liang}},\ }\bibfield  {title} {\bibinfo
  {title} {$f(q,t)$ gravity},\ }\bibfield  {journal} {\bibinfo  {journal} {Eur.
  Phys. J. C}\ }\href {https://doi.org/doi:10.1140/epjc/s10052-019-7207-4}
  {doi:10.1140/epjc/s10052-019-7207-4} ({\natexlab{a}})\BibitemShut {NoStop}%
\bibitem [{\citenamefont {Starobinsky}(1980)}]{Starobinsky:1980te}%
  \BibitemOpen
  \bibfield  {author} {\bibinfo {author} {\bibfnamefont {A.~A.}\ \bibnamefont
  {Starobinsky}},\ }\bibfield  {title} {\bibinfo {title} {{A New Type of
  Isotropic Cosmological Models Without Singularity}},\ }\href
  {https://doi.org/10.1016/0370-2693(80)90670-X} {\bibfield  {journal}
  {\bibinfo  {journal} {Phys. Lett. B}\ }\textbf {\bibinfo {volume} {91}},\
  \bibinfo {pages} {99} (\bibinfo {year} {1980})}\BibitemShut {NoStop}%
\bibitem [{\citenamefont {D'Ambrosio}\ \emph {et~al.}(2022)\citenamefont
  {D'Ambrosio}, \citenamefont {Fell}, \citenamefont {Heisenberg},\ and\
  \citenamefont {Kuhn}}]{PhysRevD.105.024042}%
  \BibitemOpen
  \bibfield  {author} {\bibinfo {author} {\bibfnamefont {F.}~\bibnamefont
  {D'Ambrosio}}, \bibinfo {author} {\bibfnamefont {S.~D.~B.}\ \bibnamefont
  {Fell}}, \bibinfo {author} {\bibfnamefont {L.}~\bibnamefont {Heisenberg}},\
  and\ \bibinfo {author} {\bibfnamefont {S.}~\bibnamefont {Kuhn}},\ }\bibfield
  {title} {\bibinfo {title} {Black holes in $f(\mathbb{Q})$ gravity},\ }\href
  {https://doi.org/10.1103/PhysRevD.105.024042} {\bibfield  {journal} {\bibinfo
   {journal} {Phys. Rev. D}\ }\textbf {\bibinfo {volume} {105}},\ \bibinfo
  {pages} {024042} (\bibinfo {year} {2022})}\BibitemShut {NoStop}%
\bibitem [{\citenamefont {Deng}\ and\ \citenamefont
  {Huang}(2017)}]{Deng:2017abh}%
  \BibitemOpen
  \bibfield  {author} {\bibinfo {author} {\bibfnamefont {G.-M.}\ \bibnamefont
  {Deng}}\ and\ \bibinfo {author} {\bibfnamefont {Y.-C.}\ \bibnamefont
  {Huang}},\ }\bibfield  {title} {\bibinfo {title} {{$Q$-$\Phi$ criticality and
  microstructure of charged AdS black holes in $f(R)$ gravity}},\ }\href
  {https://doi.org/10.1142/S0217751X17502049} {\bibfield  {journal} {\bibinfo
  {journal} {Int. J. Mod. Phys. A}\ }\textbf {\bibinfo {volume} {32}},\
  \bibinfo {pages} {1750204} (\bibinfo {year} {2017})},\ \Eprint
  {https://arxiv.org/abs/1705.04923} {arXiv:1705.04923 [gr-qc]} \BibitemShut
  {NoStop}%
\bibitem [{\citenamefont {Cai}\ and\ \citenamefont {Soh}(1999)}]{Cai:1998vy}%
  \BibitemOpen
  \bibfield  {author} {\bibinfo {author} {\bibfnamefont {R.-G.}\ \bibnamefont
  {Cai}}\ and\ \bibinfo {author} {\bibfnamefont {K.-S.}\ \bibnamefont {Soh}},\
  }\bibfield  {title} {\bibinfo {title} {{Topological black holes in the
  dimensionally continued gravity}},\ }\href
  {https://doi.org/10.1103/PhysRevD.59.044013} {\bibfield  {journal} {\bibinfo
  {journal} {Phys. Rev. D}\ }\textbf {\bibinfo {volume} {59}},\ \bibinfo
  {pages} {044013} (\bibinfo {year} {1999})},\ \Eprint
  {https://arxiv.org/abs/gr-qc/9808067} {arXiv:gr-qc/9808067} \BibitemShut
  {NoStop}%
\bibitem [{\citenamefont {Cai}(2002)}]{PhysRevD.65.084014}%
  \BibitemOpen
  \bibfield  {author} {\bibinfo {author} {\bibfnamefont {R.-G.}\ \bibnamefont
  {Cai}},\ }\bibfield  {title} {\bibinfo {title} {Gauss-bonnet black holes in
  ads spaces},\ }\href {https://doi.org/10.1103/PhysRevD.65.084014} {\bibfield
  {journal} {\bibinfo  {journal} {Phys. Rev. D}\ }\textbf {\bibinfo {volume}
  {65}},\ \bibinfo {pages} {084014} (\bibinfo {year} {2002})}\BibitemShut
  {NoStop}%
\bibitem [{\citenamefont {Zhang}\ \emph {et~al.}(2021)\citenamefont {Zhang},
  \citenamefont {Zhang}, \citenamefont {Zou},\ and\ \citenamefont
  {Yue}}]{Zhang:2021raw}%
  \BibitemOpen
  \bibfield  {author} {\bibinfo {author} {\bibfnamefont {M.}~\bibnamefont
  {Zhang}}, \bibinfo {author} {\bibfnamefont {C.-M.}\ \bibnamefont {Zhang}},
  \bibinfo {author} {\bibfnamefont {D.-C.}\ \bibnamefont {Zou}},\ and\ \bibinfo
  {author} {\bibfnamefont {R.-H.}\ \bibnamefont {Yue}},\ }\bibfield  {title}
  {\bibinfo {title} {{P \ensuremath{-} V criticality and Joule-Thomson
  expansion of Hayward-AdS black holes in 4D Einstein-Gauss-Bonnet gravity}},\
  }\href {https://doi.org/10.1016/j.nuclphysb.2021.115608} {\bibfield
  {journal} {\bibinfo  {journal} {Nucl. Phys. B}\ }\textbf {\bibinfo {volume}
  {973}},\ \bibinfo {pages} {115608} (\bibinfo {year} {2021})},\ \Eprint
  {https://arxiv.org/abs/2102.04308} {arXiv:2102.04308 [hep-th]} \BibitemShut
  {NoStop}%
\bibitem [{\citenamefont {Sharif}\ and\ \citenamefont
  {Khan}(2022)}]{Sharif:2022ccc}%
  \BibitemOpen
  \bibfield  {author} {\bibinfo {author} {\bibfnamefont {M.}~\bibnamefont
  {Sharif}}\ and\ \bibinfo {author} {\bibfnamefont {A.}~\bibnamefont {Khan}},\
  }\bibfield  {title} {\bibinfo {title} {{Thermal fluctuations, quasi-normal
  modes and phase transitions of regular black hole}},\ }\href
  {https://doi.org/10.1016/j.cjph.2022.01.002} {\bibfield  {journal} {\bibinfo
  {journal} {Chin. J. Phys.}\ }\textbf {\bibinfo {volume} {77}},\ \bibinfo
  {pages} {1885} (\bibinfo {year} {2022})}\BibitemShut {NoStop}%
\bibitem [{\citenamefont {Ezroura}\ \emph {et~al.}(2022)\citenamefont
  {Ezroura}, \citenamefont {Larsen}, \citenamefont {Liu},\ and\ \citenamefont
  {Zeng}}]{Ezroura:2021vrt}%
  \BibitemOpen
  \bibfield  {author} {\bibinfo {author} {\bibfnamefont {N.}~\bibnamefont
  {Ezroura}}, \bibinfo {author} {\bibfnamefont {F.}~\bibnamefont {Larsen}},
  \bibinfo {author} {\bibfnamefont {Z.}~\bibnamefont {Liu}},\ and\ \bibinfo
  {author} {\bibfnamefont {Y.}~\bibnamefont {Zeng}},\ }\bibfield  {title}
  {\bibinfo {title} {{The phase diagram of BPS black holes in AdS$_{5}$}},\
  }\href {https://doi.org/10.1007/JHEP09(2022)033} {\bibfield  {journal}
  {\bibinfo  {journal} {JHEP}\ }\textbf {\bibinfo {volume} {09}},\ \bibinfo
  {pages} {033}},\ \Eprint {https://arxiv.org/abs/2108.11542} {arXiv:2108.11542
  [hep-th]} \BibitemShut {NoStop}%
\bibitem [{\citenamefont {Belhaj}\ \emph {et~al.}(2021)\citenamefont {Belhaj},
  \citenamefont {El~Balali}, \citenamefont {El~Hadri}, \citenamefont
  {Hassouni},\ and\ \citenamefont {Torrente-Lujan}}]{Belhaj:2020pnh}%
  \BibitemOpen
  \bibfield  {author} {\bibinfo {author} {\bibfnamefont {A.}~\bibnamefont
  {Belhaj}}, \bibinfo {author} {\bibfnamefont {A.}~\bibnamefont {El~Balali}},
  \bibinfo {author} {\bibfnamefont {W.}~\bibnamefont {El~Hadri}}, \bibinfo
  {author} {\bibfnamefont {Y.}~\bibnamefont {Hassouni}},\ and\ \bibinfo
  {author} {\bibfnamefont {E.}~\bibnamefont {Torrente-Lujan}},\ }\bibfield
  {title} {\bibinfo {title} {{Phase transition and shadow behaviors of
  quintessential black holes in M-theory/superstring inspired models}},\ }\href
  {https://doi.org/10.1142/S0217751X21500573} {\bibfield  {journal} {\bibinfo
  {journal} {Int. J. Mod. Phys. A}\ }\textbf {\bibinfo {volume} {36}},\
  \bibinfo {pages} {2150057} (\bibinfo {year} {2021})},\ \Eprint
  {https://arxiv.org/abs/2004.10647} {arXiv:2004.10647 [hep-th]} \BibitemShut
  {NoStop}%
\bibitem [{\citenamefont {Su}\ \emph {et~al.}(2020)\citenamefont {Su},
  \citenamefont {Wang},\ and\ \citenamefont {Li}}]{Su:2019gby}%
  \BibitemOpen
  \bibfield  {author} {\bibinfo {author} {\bibfnamefont {B.-Y.}\ \bibnamefont
  {Su}}, \bibinfo {author} {\bibfnamefont {Y.-Y.}\ \bibnamefont {Wang}},\ and\
  \bibinfo {author} {\bibfnamefont {N.}~\bibnamefont {Li}},\ }\bibfield
  {title} {\bibinfo {title} {{The Hawking\textendash{}Page phase transitions in
  the extended phase space in the Gauss\textendash{}Bonnet gravity}},\ }\href
  {https://doi.org/10.1140/epjc/s10052-020-7870-5} {\bibfield  {journal}
  {\bibinfo  {journal} {Eur. Phys. J. C}\ }\textbf {\bibinfo {volume} {80}},\
  \bibinfo {pages} {305} (\bibinfo {year} {2020})},\ \Eprint
  {https://arxiv.org/abs/1905.07155} {arXiv:1905.07155 [gr-qc]} \BibitemShut
  {NoStop}%
\bibitem [{\citenamefont {Ruppeiner}()}]{ruppeiner1995riemannian}%
  \BibitemOpen
  \bibfield  {author} {\bibinfo {author} {\bibfnamefont {G.}~\bibnamefont
  {Ruppeiner}},\ }\bibfield  {title} {\bibinfo {title} {Riemannian geometry in
  thermodynamic fluctuation theory},\ }\bibfield  {journal} {\bibinfo
  {journal} {Reviews of Modern Physics}\ }\href
  {https://doi.org/https://doi.org/10.1103/RevModPhys.67.605}
  {https://doi.org/10.1103/RevModPhys.67.605}\BibitemShut {NoStop}%
\bibitem [{\citenamefont {Weinhold}()}]{weinhold1975metric}%
  \BibitemOpen
  \bibfield  {author} {\bibinfo {author} {\bibfnamefont {F.}~\bibnamefont
  {Weinhold}},\ }\bibfield  {title} {\bibinfo {title} {Metric geometry of
  equilibrium thermodynamics},\ }\bibfield  {journal} {\bibinfo  {journal} {The
  Journal of Chemical Physics}\ }\textbf {\bibinfo {volume} {63}},\ \href
  {https://doi.org/10.1063/1.431635} {10.1063/1.431635}\BibitemShut {NoStop}%
\bibitem [{\citenamefont {Salamon}\ \emph {et~al.}()\citenamefont {Salamon},
  \citenamefont {Nulton},\ and\ \citenamefont {Ihrig}}]{salamon1984relation}%
  \BibitemOpen
  \bibfield  {author} {\bibinfo {author} {\bibfnamefont {P.}~\bibnamefont
  {Salamon}}, \bibinfo {author} {\bibfnamefont {J.}~\bibnamefont {Nulton}},\
  and\ \bibinfo {author} {\bibfnamefont {E.}~\bibnamefont {Ihrig}},\ }\bibfield
   {title} {\bibinfo {title} {On the relation between entropy and energy
  versions of thermodynamic length},\ }\bibfield  {journal} {\bibinfo
  {journal} {The Journal of chemical physics}\ }\textbf {\bibinfo {volume}
  {80}},\ \href {https://doi.org/https://doi.org/10.1063/1.446467}
  {https://doi.org/10.1063/1.446467}\BibitemShut {NoStop}%
\bibitem [{\citenamefont {Mruga{\l}a}()}]{mrugala1984equivalence}%
  \BibitemOpen
  \bibfield  {author} {\bibinfo {author} {\bibfnamefont {R.}~\bibnamefont
  {Mruga{\l}a}},\ }\bibfield  {title} {\bibinfo {title} {On equivalence of two
  metrics in classical thermodynamics},\ }\bibfield  {journal} {\bibinfo
  {journal} {Physica A: Statistical Mechanics and its Applications}\ }\href
  {https://doi.org/https://doi.org/10.1016/0378-4371(84)90074-8}
  {https://doi.org/10.1016/0378-4371(84)90074-8}\BibitemShut {NoStop}%
\bibitem [{\citenamefont {Xu}\ \emph {et~al.}({\natexlab{b}})\citenamefont
  {Xu}, \citenamefont {Wu},\ and\ \citenamefont {Yang}}]{xu2020ruppeiner}%
  \BibitemOpen
  \bibfield  {author} {\bibinfo {author} {\bibfnamefont {Z.-M.}\ \bibnamefont
  {Xu}}, \bibinfo {author} {\bibfnamefont {B.}~\bibnamefont {Wu}},\ and\
  \bibinfo {author} {\bibfnamefont {W.-L.}\ \bibnamefont {Yang}},\ }\bibfield
  {title} {\bibinfo {title} {Ruppeiner thermodynamic geometry for the
  schwarzschild-ads black hole},\ }\bibfield  {journal} {\bibinfo  {journal}
  {Physical Review D}\ }\textbf {\bibinfo {volume} {101}},\ \href
  {https://doi.org/https://doi.org/10.1103/PhysRevD.101.024018}
  {https://doi.org/10.1103/PhysRevD.101.024018} ({\natexlab{b}})\BibitemShut
  {NoStop}%
\bibitem [{\citenamefont {Ghosh}\ and\ \citenamefont
  {Bhamidipati}()}]{ghosh2020thermodynamic}%
  \BibitemOpen
  \bibfield  {author} {\bibinfo {author} {\bibfnamefont {A.}~\bibnamefont
  {Ghosh}}\ and\ \bibinfo {author} {\bibfnamefont {C.}~\bibnamefont
  {Bhamidipati}},\ }\bibfield  {title} {\bibinfo {title} {Thermodynamic
  geometry for charged gauss-bonnet black holes in ads spacetimes},\ }\bibfield
   {journal} {\bibinfo  {journal} {Physical Review D}\ }\href
  {https://doi.org/https://doi.org/10.1103/PhysRevD.101.046005}
  {https://doi.org/10.1103/PhysRevD.101.046005}\BibitemShut {NoStop}%
\bibitem [{\citenamefont {Caldarelli}\ \emph {et~al.}(2000)\citenamefont
  {Caldarelli}, \citenamefont {Cognola},\ and\ \citenamefont
  {Klemm}}]{Caldarelli:1999xj}%
  \BibitemOpen
  \bibfield  {author} {\bibinfo {author} {\bibfnamefont {M.~M.}\ \bibnamefont
  {Caldarelli}}, \bibinfo {author} {\bibfnamefont {G.}~\bibnamefont
  {Cognola}},\ and\ \bibinfo {author} {\bibfnamefont {D.}~\bibnamefont
  {Klemm}},\ }\bibfield  {title} {\bibinfo {title} {{Thermodynamics of
  Kerr-Newman-AdS black holes and conformal field theories}},\ }\href
  {https://doi.org/10.1088/0264-9381/17/2/310} {\bibfield  {journal} {\bibinfo
  {journal} {Class. Quant. Grav.}\ }\textbf {\bibinfo {volume} {17}},\ \bibinfo
  {pages} {399} (\bibinfo {year} {2000})},\ \Eprint
  {https://arxiv.org/abs/hep-th/9908022} {arXiv:hep-th/9908022} \BibitemShut
  {NoStop}%
\bibitem [{\citenamefont {Cvetic}\ \emph {et~al.}(2002)\citenamefont {Cvetic},
  \citenamefont {Nojiri},\ and\ \citenamefont {Odintsov}}]{Cvetic:2001bk}%
  \BibitemOpen
  \bibfield  {author} {\bibinfo {author} {\bibfnamefont {M.}~\bibnamefont
  {Cvetic}}, \bibinfo {author} {\bibfnamefont {S.}~\bibnamefont {Nojiri}},\
  and\ \bibinfo {author} {\bibfnamefont {S.~D.}\ \bibnamefont {Odintsov}},\
  }\bibfield  {title} {\bibinfo {title} {{Black hole thermodynamics and
  negative entropy in de Sitter and anti-de Sitter Einstein-Gauss-Bonnet
  gravity}},\ }\href {https://doi.org/10.1016/S0550-3213(02)00075-5} {\bibfield
   {journal} {\bibinfo  {journal} {Nucl. Phys. B}\ }\textbf {\bibinfo {volume}
  {628}},\ \bibinfo {pages} {295} (\bibinfo {year} {2002})},\ \Eprint
  {https://arxiv.org/abs/hep-th/0112045} {arXiv:hep-th/0112045} \BibitemShut
  {NoStop}%
\bibitem [{\citenamefont {Rinaldi}(2012)}]{Rinaldi:2012vy}%
  \BibitemOpen
  \bibfield  {author} {\bibinfo {author} {\bibfnamefont {M.}~\bibnamefont
  {Rinaldi}},\ }\bibfield  {title} {\bibinfo {title} {{Black holes with
  non-minimal derivative coupling}},\ }\href
  {https://doi.org/10.1103/PhysRevD.86.084048} {\bibfield  {journal} {\bibinfo
  {journal} {Phys. Rev. D}\ }\textbf {\bibinfo {volume} {86}},\ \bibinfo
  {pages} {084048} (\bibinfo {year} {2012})},\ \Eprint
  {https://arxiv.org/abs/1208.0103} {arXiv:1208.0103 [gr-qc]} \BibitemShut
  {NoStop}%
\bibitem [{\citenamefont {Chen}\ \emph {et~al.}(2023)\citenamefont {Chen},
  \citenamefont {Chen}, \citenamefont {Ishibashi},\ and\ \citenamefont
  {Ohta}}]{Chen:2023pcv}%
  \BibitemOpen
  \bibfield  {author} {\bibinfo {author} {\bibfnamefont {C.-M.}\ \bibnamefont
  {Chen}}, \bibinfo {author} {\bibfnamefont {Y.}~\bibnamefont {Chen}}, \bibinfo
  {author} {\bibfnamefont {A.}~\bibnamefont {Ishibashi}},\ and\ \bibinfo
  {author} {\bibfnamefont {N.}~\bibnamefont {Ohta}},\ }\bibfield  {title}
  {\bibinfo {title} {{Phase Structure of Quantum Improved
  Schwarzschild-(Anti)de Sitter Black Holes}},\ }\href@noop {} {\  (\bibinfo
  {year} {2023})},\ \Eprint {https://arxiv.org/abs/2303.04304}
  {arXiv:2303.04304 [hep-th]} \BibitemShut {NoStop}%
\bibitem [{\citenamefont {Karmakar}\ \emph {et~al.}(2023)\citenamefont
  {Karmakar}, \citenamefont {Gogoi},\ and\ \citenamefont
  {Goswami}}]{Karmakar:2023mhs}%
  \BibitemOpen
  \bibfield  {author} {\bibinfo {author} {\bibfnamefont {R.}~\bibnamefont
  {Karmakar}}, \bibinfo {author} {\bibfnamefont {D.~J.}\ \bibnamefont
  {Gogoi}},\ and\ \bibinfo {author} {\bibfnamefont {U.~D.}\ \bibnamefont
  {Goswami}},\ }\bibfield  {title} {\bibinfo {title} {{Thermodynamics and
  Shadows of GUP-corrected Black Holes with Topological Defects in Bumblebee
  Gravity}},\ }\href@noop {} {\  (\bibinfo {year} {2023})},\ \Eprint
  {https://arxiv.org/abs/2303.00297} {arXiv:2303.00297 [gr-qc]} \BibitemShut
  {NoStop}%
\bibitem [{\citenamefont {Jafarzade}\ \emph {et~al.}(2023)\citenamefont
  {Jafarzade}, \citenamefont {Rezaei},\ and\ \citenamefont
  {Hendi}}]{Jafarzade:2023yof}%
  \BibitemOpen
  \bibfield  {author} {\bibinfo {author} {\bibfnamefont {K.}~\bibnamefont
  {Jafarzade}}, \bibinfo {author} {\bibfnamefont {E.}~\bibnamefont {Rezaei}},\
  and\ \bibinfo {author} {\bibfnamefont {S.~H.}\ \bibnamefont {Hendi}},\
  }\bibfield  {title} {\bibinfo {title} {{Rotating Lifshitz-like black holes in
  F(R) gravity}},\ }\href@noop {} {\  (\bibinfo {year} {2023})},\ \Eprint
  {https://arxiv.org/abs/2302.03500} {arXiv:2302.03500 [hep-th]} \BibitemShut
  {NoStop}%
\bibitem [{\citenamefont {Czinner}\ and\ \citenamefont
  {Iguchi}(2016)}]{Czinner:2015eyk}%
  \BibitemOpen
  \bibfield  {author} {\bibinfo {author} {\bibfnamefont {V.~G.}\ \bibnamefont
  {Czinner}}\ and\ \bibinfo {author} {\bibfnamefont {H.}~\bibnamefont
  {Iguchi}},\ }\bibfield  {title} {\bibinfo {title} {{R\'enyi Entropy and the
  Thermodynamic Stability of Black Holes}},\ }\href
  {https://doi.org/10.1016/j.physletb.2015.11.061} {\bibfield  {journal}
  {\bibinfo  {journal} {Phys. Lett. B}\ }\textbf {\bibinfo {volume} {752}},\
  \bibinfo {pages} {306} (\bibinfo {year} {2016})},\ \Eprint
  {https://arxiv.org/abs/1511.06963} {arXiv:1511.06963 [gr-qc]} \BibitemShut
  {NoStop}%
\bibitem [{\citenamefont {Altamirano}\ \emph {et~al.}(2014)\citenamefont
  {Altamirano}, \citenamefont {Kubiznak}, \citenamefont {Mann},\ and\
  \citenamefont {Sherkatghanad}}]{Altamirano:2014tva}%
  \BibitemOpen
  \bibfield  {author} {\bibinfo {author} {\bibfnamefont {N.}~\bibnamefont
  {Altamirano}}, \bibinfo {author} {\bibfnamefont {D.}~\bibnamefont
  {Kubiznak}}, \bibinfo {author} {\bibfnamefont {R.~B.}\ \bibnamefont {Mann}},\
  and\ \bibinfo {author} {\bibfnamefont {Z.}~\bibnamefont {Sherkatghanad}},\
  }\bibfield  {title} {\bibinfo {title} {{Thermodynamics of rotating black
  holes and black rings: phase transitions and thermodynamic volume}},\ }\href
  {https://doi.org/10.3390/galaxies2010089} {\bibfield  {journal} {\bibinfo
  {journal} {Galaxies}\ }\textbf {\bibinfo {volume} {2}},\ \bibinfo {pages}
  {89} (\bibinfo {year} {2014})},\ \Eprint {https://arxiv.org/abs/1401.2586}
  {arXiv:1401.2586 [hep-th]} \BibitemShut {NoStop}%
\bibitem [{\citenamefont {Yu}\ \emph {et~al.}(2023)\citenamefont {Yu},
  \citenamefont {Xu},\ and\ \citenamefont {Tao}}]{yu2023thermodynamics}%
  \BibitemOpen
  \bibfield  {author} {\bibinfo {author} {\bibfnamefont {Q.}~\bibnamefont
  {Yu}}, \bibinfo {author} {\bibfnamefont {Q.}~\bibnamefont {Xu}},\ and\
  \bibinfo {author} {\bibfnamefont {J.}~\bibnamefont {Tao}},\ }\bibfield
  {title} {\bibinfo {title} {Thermodynamics and microstructures of
  euler-heisenberg black hole in a cavity}\ }\href {https://doi.org/arXiv
  preprint arXiv:2302.09821} {arXiv preprint arXiv:2302.09821} (\bibinfo {year}
  {2023})\BibitemShut {NoStop}%
\end{thebibliography}
\end{document}